\documentclass[pre, aps,twocolumn,floatfix,superscriptaddress, nofootinbib]{revtex4-1}
\usepackage{amsmath,amssymb,graphicx,cases}
\usepackage{epsfig}
\usepackage{booktabs}
\usepackage{textcomp}
\usepackage{epstopdf}
\usepackage{graphicx}
\usepackage{float}
\usepackage{braket}
\usepackage[normalem]{ulem}
\usepackage[utf8]{inputenc}
\usepackage{enumitem} 

\usepackage[normalem]{ulem}
\usepackage{dsfont}
\newcommand{\stkout}[1]{\ifmmode\text{\sout{\ensuremath{#1}}}\else\sout{#1}\fi}

\usepackage{fixmath}

\usepackage{color}
\definecolor{Blue}{rgb}{0.00, 0.00, 1.00}
\definecolor{Red}{rgb}{1.00, 0.00, 0.00}
\definecolor{Green}{rgb}{0.00, 0.60, 0.00}

\newcommand{\nn}{\nonumber}
\newcommand{\be}{\begin{equation}}
\newcommand{\ee}{\end{equation}}
\newcommand{\bea}{\begin{eqnarray}}
\newcommand{\eea}{\end{eqnarray}}

\usepackage{multirow}
\usepackage{float}
\usepackage{caption, subcaption}

\usepackage[colorlinks=true, urlcolor=blue, anchorcolor=blue, citecolor=blue,filecolor=blue,linkcolor=blue,menucolor=blue]{hyperref}

\begin{document}
\title{Large deviations in statistics of the local time and occupation time for a run and tumble particle}

\author{Soheli Mukherjee}
\email{soheli.mukherjee2@gmail.com}

\affiliation{Department of Environmental Physics,
  Blaustein Institutes for Desert Research, Ben-Gurion University of
  the Negev, Sede Boqer Campus, 8499000, Israel}

\author{Pierre Le Doussal}
\address{Laboratoire de Physique de l'Ecole Normale Sup\'erieure, CNRS, ENS and PSL Universit\'e, Sorbonne Universit\'e, Universit\'e Paris Cit\'e,
24 rue Lhomond, 75005 Paris, France}

\author{Naftali R. Smith}
\email{naftalismith@gmail.com}
\affiliation{Department of Environmental Physics,
  Blaustein Institutes for Desert Research, Ben-Gurion University of
  the Negev, Sede Boqer Campus, 8499000, Israel}

\begin{abstract}
We investigate the statistics of the local time $\mathcal{T} = \int_0^T  \delta(x(t)) dt$ that a run and tumble particle (RTP) $x(t)$ in one dimension spends at the origin, with or without an external drift.
By relating the local time to the number of times the RTP crosses the origin, we find that the local time distribution $P(\mathcal{T})$ satisfies the large deviation principle $P(\mathcal{T}) \sim \, e^{-T \, I(\mathcal{T} / T)}
$ in the  large observation time limit $T \to \infty$.
Remarkably, we find that in presence of drift the rate function $I(\rho)$ is nonanalytic: We interpret its singularity as a dynamical phase transition of first order.
We then extend these results by studying the statistics of the amount of time $\mathcal{R}$ that the RTP spends inside a finite interval (i.e., the occupation time), with qualitatively similar results. In particular, this yields the long-time decay rate of the probability $P(\mathcal{R} = T)$ that the particle does not exit the interval up to time $T$.
We find that the conditional endpoint distribution exhibits an interesting change of behavior from unimodal to bimodal as a function of the size of the interval. To study the occupation time statistics, we extend the Donsker-Varadhan large-deviation formalism to the case of RTPs, for general dynamical observables and possibly in the presence of an external potential. 
\end{abstract}

\maketitle

\section{Introduction}

The local time of a particle refers to the time spent by the particle in the neighbourhood of a given point in space during an observation time $T$. It is a very important quantity and has useful applications in various scientific fields. For example, in the context of chemical and biological reactions, the reaction rate is calculated by the local time that the reactant spends in the vicinity of the receptor \cite{WF1973, BCKMO2005, Doi1975}.  
The local time is also proportional to the concentration of monomers in a polymer, in bacterial chemotaxis which is affected by the local time spent by a bacteria at a point \cite{Koshland1980}. 
Another interesting quantity to consider is the occupation time, which is the time a particle spends in a given region of space. The local time is closely related to the limiting case of the occupation time in the limit of zero domain size. 

Fluctuations of the local time and occupation time have been studied extensively for different processes:  the Ornstein-Uhlenbeck process \cite{KK2021}, continuous time random walk \cite{CTB2010}, theory and experiments of blinking quantum dots \cite{SHB2009}, study of diffusion in porous rocks as modelled by diffusion in a random potential landscape \cite{MC2002, SMC2006}, diffusion on graphs \cite{CDM2002} and diffusion under stochastic resetting \cite{PCRK2019}. From a theoretical point of view the study of the local time or the occupation time is important for stochastic processes as it provides information about the spatiotemporal correlations of the particle’s trajectory.

While the local time and occupation time have been studied extensively for Brownian particles \cite{CDM2002, PCRK2019, BMR23, SmithMeerson24}, there are very few existing results in the biologically-relevant context of active matter \cite{Schweitzer2003, RBELS2012, MJRLPRS2013,  FGGVWW2015, GKKRST23}.
The time spent by drug molecules in the vicinity of their target is of crucial importance for their efficiency, and it has been argued that the activity of drug molecules may enhance their ability to reach the target \cite{Santiago2018, GXGG2020}.
In Ref.~\cite{SK21} the local time for a run and tumble particle (RTP) in one dimension was studied. They obtained an exact result for the distribution of the local time, in the absence of an external drift. They then extracted the behavior in the regime of typical fluctuations at long times. The large-deviation regime was not studied in detail. 

In general, large deviations \cite{DZ1998, Hollander2000, Hugo2009} are of great interest because they can have dramatic consequences (e.g., earthquakes, stock-market crashes etc).
They are also of fundamental interest in statistical physics because the large deviation functions of appropriate observable acts as a substitute of free energy in non-equilibrium systems. 
One of the standard types of problems studied in large-deviation theory is to calculate the distribution of ``dynamical observables", which are given by time averages of the stochastic process under study.
One general form of such dynamical observables is given by
\be \label{observable}
 \mathcal{A} =  \int_0^T \, f(x(t)) \,\, dt,
\ee
where $x(t)$ is a stochastic process (particular examples will be given below), and $f(x)$ is some function.
In the limit $T\to\infty$, 
assuming the process is ergodic,
 the time average and ensemble average coincide, and therefore, with probability one, $\mathcal{A}/T$ will equal its corresponding ensemble-average value. 
However, if $T$ is large but finite, there will be fluctuations away from this expected value which are of interest.

 For finite observation time $T$
its probability density function (PDF), $P(\mathcal{A})$, 
depends on details such  as
the initial position $x_0=x(0)$ of the 
Brownian motion. However in the long $T$ limit, 
it satisfies a  universal large deviation principle (LDP) \cite{Hugo2009}  
\be 
P(\mathcal{A}) \sim \, e^{-T \, I(\mathcal{A} / T)} \label{LDP0},
\ee
where $I(\rho)$  is the rate function which is defined as 
\be
I(\rho) = - \lim_{T \to \infty} \frac{1}{T} \ln P (\mathcal{A} = \rho T). 
\ee
According to G{\"a}rtner-Ellis theorem, the rate function $I(\rho)$ can be calculated via the Legendre-Fenchel transform of the scaled cumulant generating function (SCGF) \cite{Hugo2009}, which is defined as 
\bea \label{eqscgf}
\lambda(k) = \lim_{T \to \infty} \frac{1}{T} \,\, \ln \langle  e^{ k \, \mathcal{A}} \rangle \, ,
\eea
 which is by construction a convex function for real $k$.

Using Eq. \eqref{LDP0}, the expectation value in Eq. \eqref{eqscgf} is dominated at large $T$ by a saddle point, leading to
 \be 
 \lambda(k) = \max_{\rho \geq 0} ( k \, \rho - I(\rho) ).
 \ee 
The rate function is then expressed as the Legendre-Fenchel transform of $\lambda(k)$
\be
\label{LegendreFenchel}
I(\rho) = \sup_{k} \lbrace  k \, \rho - \lambda(k) \rbrace,
\ee
provided that $\lambda(k)$ exists and is differentiable. If $\lambda(k)$ is not differentiable, then  $I(\rho)$ is replaced by its convex envelope \cite{Hugo2009}.

If $x(t)$ is a Brownian motion (possibly with external forces) then one can use the well-established Donsker-Varadhan (DV) theory to calculate $\lambda(k)$ \cite{Hugo2018}.
This theory states that (under certain conditions) $\lambda(k)$ is given by the largest eigenvalue of a ($k$-dependent) linear operator that is related to the generator of the dynamics of the position distribution of the particle.

In particular, the DV formalism was used successfully in Refs.~\cite{NT2018, NT2018DPT} to calculate large deviations of the occupation time of a drifted Brownian particle in a finite interval. Remarkably, they found that (for nonzero drift) the associated rate function contains a singularity. Such singularities are usually interpreted as dynamical phase transitions (DPT), in analogy with equilibrium statistical physics \cite{exclusion, glass, kafri, exclusion1, singularities, baek, baek1, NT2018, NT2018DPT, MukherjeeSmith23, Smith22Chaos, MukherjeeSmithConvexhull}.
 Large deviations of the local time for a discrete model of a RTP on a lattice were studied in \cite{MBE19}. In the presence of an external drift, a DPT was uncovered, which is qualitatively similar to the DPT from the Brownian case \cite{NT2018, NT2018DPT}.

 In this paper we consider a continuum model for a RTP in one dimension.
The two observables under study in this work are particular cases of Eq.~\eqref{observable}: local time and occupation time, defined as
\begin{eqnarray} \label{localtime}
    \text{Local time}: \qquad \mathcal{T} &=& \int_0^T \, \delta(x(t))  dt, \\
     \text{Occupation time}: \qquad \mathcal{R} &=& \int_0^T \, \mathds{1}_{[-a,a]}(x(t))  dt,\label{occupation} 
\end{eqnarray}
respectively, where $\delta(x)$ is the Dirac delta function  and $\mathds{1}_{[-a,a]}(x)$ is the indicator function:
\begin{equation}
      \mathds{1}_{[-a,a]}(x) =  \begin{cases}
    1 &  \text{for }  -a \leq x \leq a, \\[1mm]
    0  & \text{otherwise}. \\
    \end{cases}
\end{equation}
The mean values of these observables scale, at long times, as $\sim \sqrt{T}$  in the absence of a drift, and as $\sim T^0$ if a drift is present. We study here the large deviations regime that describes fluctuations of order $\sim T$ of the local time and occupation time for RTP's, with or without external drifts, i.e., we focus here on the right tails of  the distributions of $\mathcal{T}$ and $\mathcal{R}$.
 In particular, the probability that $\mathcal{R} = T$ is the survival probability \cite{Weiss84, WMLW87, MPW92, MalakarEtAl18,MLMS20, BMS21, NS24} of the particle inside the interval.
 In fact, in doing so, we will also extend the general theory of large deviations of dynamical observables \eqref{observable} (i.e., the DV theory) to the case in which $x(t)$ describes the motion of an RTP in one dimension.
As in the case of the Brownian motion, here too we find that, in the case of nonzero drift, the rate functions describing the local time or occupation time distribution contain singularities which we interpret as first-order DPTs.

The remainder of the paper is organized as follows. In Section \ref{BM with drift}, for completeness, we study large deviations of the local time for Brownian motion. The analysis is very similar to that of Refs.~\cite{NT2018, NT2018DPT} in which the occupation time in an interval was studied. In Section \ref{RTP local time from crossing} we study the local time for an RTP by exploiting a simple connection between the local time and the number of times that the particle crossed the origin.
In Section \ref{DV for RTP} we begin by extending DV theory to RTPs for general observables \eqref{observable}, and we then apply this theory to study the occupation time.
In Section \ref{conclusion} we summarize and discuss our main findings.

\section{Local time for Brownian motion with drift} \label{BM with drift}

We consider a Brownian motion $x(t)$ with a drift, as described by the equation of motion
\be
\dot{x}=\mu + \sqrt{2D}\,\xi(t) \, .
\ee
 Here $\xi(t)$ is a standard Gaussian white noise with  $\left\langle \xi\left(t\right)\right\rangle =0$ and 
 $\left\langle \xi\left(t\right)\xi\left(t'\right)\right\rangle =\delta\left(t-t'\right)$. $\mu \geq 0$  is the external drift.
 We are interested in the distribution of the local time $\mathcal{T}$,
which is the time that a particle spends near the origin $x=0$ during the time of observation $T$.
It is defined as
\begin{eqnarray} \label{localtime}
    \mathcal{T} = \int_0^T \, \delta(x(t)) \,\, dt,
\end{eqnarray}
so that $0 \leq \mathcal{T} < \infty$. 
We prove that the fluctuations of the local time $\mathcal{T}$ in the large $T$ limit satisfies the LDP 
\be 
P(\mathcal{T}) \sim \, e^{-T \, I(\mathcal{T} / T)} \, .
\label{LDP}
\ee 

 According to the DV formalism, the tilted generator of the process can be written as
\bea
\mathcal{L}_{k} = \mathcal{L} + k\;\delta(x),
\eea
where 
\be
\mathcal{L} = D\;\partial_{x}^{2} + \mu \,  \partial_{x},
\ee
is the Hermitian conjugate of the Fokker-Planck generator of the drifted Brownian motion
\be
\partial_{t} \,p\left(x,t\right)=\mathcal{L}^{\dagger}\, p\left(x,t\right) \, .
\ee
Here $p(x,t)$ is the probability for the particle to be at position $x$ at time $t$. According to the DV formalism, the SCGF $\lambda(k)$ of the local time distribution is given by the largest eigenvalue of $\mathcal{L}_k$ under appropriate boundary conditions.

 The operator $\mathcal{L}_{k}$  is not Hermitian, and it is therefore convenient to define its symmetrized version. This is a standard procedure, \cite{NT2018, NT2018DPT, Hugo2018}, and it leads to the definition of the symmetrized (hermitian) operator
\be
\mathcal{H} = D \, \partial_{x}^{2} - \frac{\mu^2}{4 D} + k\,\delta(x),\ee
 whose eigenvalues coincide with those of $\mathcal{L}_k$.
The eigenvalue equation
\be
\mathcal{H} \,\psi(x) = \Tilde{\lambda} (k, \mu) \, \psi(x),
\ee
may be interpreted, up to an overall minus sign, as the time-independent
Schr\"{o}dinger equation of a quantum particle inside a Delta function potential along with a shift
\be
V_k (x) = \frac{\mu^2}{4 \, D} - k \, \delta(x) \, .
\ee
 Within this interpretation, the particle's ``energy" is given by $-\Tilde{\lambda}$. We are interested in finding the largest eigenvalue of $\mathcal{H}$, which corresponds to the ground-state energy for the potential $V_k$.
 The drift
 $\mu$ affects the potential $V_k$ by the addition of a constant term. One therefore finds that
\bea \label{groundstate BMdrift}
\Tilde{\lambda}(k, \mu) = \Tilde{\lambda} (k, 0) - \frac{\mu^2}{4 D},
\eea
where $\tilde{\lambda} (k, 0)$ is the negative of the ground state energy of a quantum particle inside a delta function potential in the absence of the constant term $-\mu^2 / (4D)$. 
 Finding the ground-state energy of a quantum particle in a delta-function potential is a standard basic exercise in quantum mechanics. The result yields
\be \label{groundstate BM}
\Tilde{\lambda}(k, 0)=\frac{k^{2}}{4D} \, ,
\ee
and the corresponding wave function is $ \psi(x)  = \sqrt{ \beta} \, e^{- \beta \, |x|}$, where  $\beta = \sqrt{\Tilde{\lambda}(k,0)/D}$.

In presence of a nonzero drift $\mu \ne 0$,  the solution \eqref{groundstate BMdrift} cannot give the true SCGF $\lambda(k)$ 
at all $k$, because it
becomes negative for $k \leq \mu$. 
 On the other hand, the true SCGF  $\lambda(k)$ must be nonnegative, since it is  a convex function of $k$ with $\lambda(0)=0$ and 
$\lambda'(0)=\lim_{T\to\infty}\left\langle \mathcal{T}/T\right\rangle =0$.
To resolve this issue, we must note that $\lambda=0$ is also an eigenvalue of $\mathcal{L}_k$ (for further technical details, we refer the reader to Refs.~\cite{NT2018,NT2018DPT} where a very similar analysis was performed in the context of the occupation time for a drifted Brownian motion).

Hence the actual  SCGF  is given by the maximum of these two eigenvalues, i.e., 
\be 
\lambda{(k)}=\max\left\{ 0,\tilde{\lambda}\left(k,\mu\right)\right\} =\begin{cases}
0\,, & k\leq k_{c}\,,\\[2mm]
\frac{k^{2}-\mu^{2}}{4D}\,, & k>k_{c}\,,
\end{cases}
\ee
where $k_c=\mu$. 
The Legendre-Fenchel transform \eqref{LegendreFenchel} then yields the rate function
\be 
I(\rho) =
\begin{cases}
\rho \, \mu\, , & \rho \leq \rho_c\, ,\\[2mm]
D \, \rho^2 + \frac{\mu^2}{4D}\, , & \rho >  \rho_c \, ,
\end{cases}
\ee
 where $\rho_c=\mu / (2D)$.

\begin{figure*}
\centering
\includegraphics[width=0.48\linewidth]{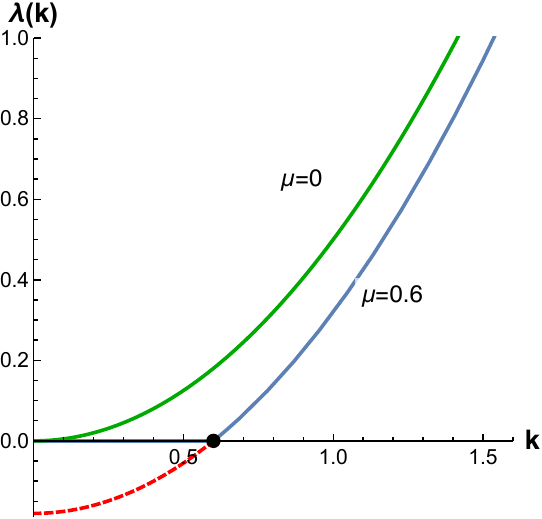}
\hspace{1mm}
\includegraphics[width=0.48\linewidth]{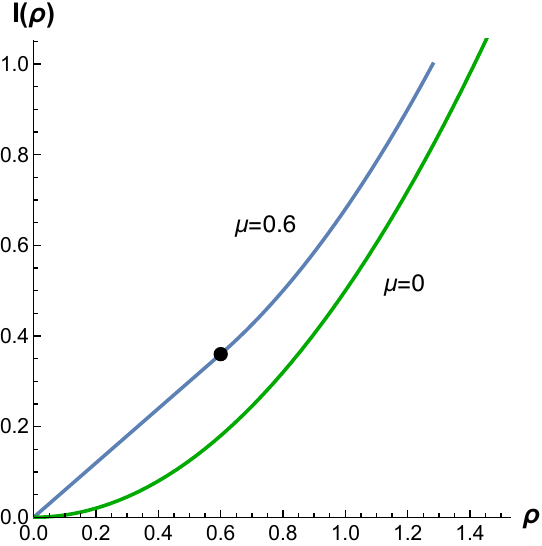}
    \caption{  Plots of the SCGF $\lambda(k)$ and of the rate function $I(\rho)$ for the local time of a Brownian particle without a drift and with a drift (here we consider $\mu = 0.6$) as depicted by the green lines and solid blue lines respectively. The solid circle denotes the first order critical point where the first derivative of $\lambda(k)$ shows a jump discontinuity, and the red dashed lines denote the unphysical solution for $\lambda(k)$.}
    \label{LDFs for brownian}
\end{figure*}

The SCGFs and the rate functions for both $\mu=0$  and $\mu \neq 0$ are shown in Fig.~\ref{LDFs for brownian}. For $\mu=0$, both the SCGF and the rate function are analytic, i.e., they contain no singularities  for $k > 0$ and $\rho >0 $. In presence of a drift, the SCGF shows a singularity at $k=k_c$: The first derivative of $\lambda(k)$ shows a jump discontinuity. This singularity is interpreted as a dynamical phase transition (DPT) of first order at $k = k_c$.  The singularity of $I(\rho)$ at the critical point $\rho=\rho_c$ is weaker: Its second derivative jumps.

 The physical mechanism behind this DPT is as follows. In the supercritical regime $\rho > \rho_c$, the particle spends the entire duration of the dynamics $t \in [0,T]$ in the vicinity of the origin $x=0$. It corresponds to the existence of a bound state in the equivalent quantum mechanical problem.
However, the subcritical regime, $0 < \rho  < \rho_c$, is a regime of temporal ``coexistence". The particle stays localized in the vicinity of the origin up to time $t \simeq \rho T / \rho_c$, and subsequently gets driven to infinity by the drift term.  The transition can also be seen as an unbinding transition
in the associated non-hermitian quantum mechanical problem \cite{HatanoNelson96}.

In the next two sections, we will study local time and occupation time for an (active) RTP. As we will see, some of the qualitative behaviors will be similar to the case of Brownian motion. In particular, we will show that, in presence of a drift, there is a first-order DPT whose physical origin is temporal coexistence between  localized and driven states.

\section{Local time for RTP with drift} \label{RTP local time from crossing}

 We now want to calculate the fluctuations of the local time defined in Eq.~\eqref{localtime} for a run and tumble particle (RTP) \cite{TC08}. 
 In the absence of an external drift, the RTP model describes a particle that propels itself at a constant speed, and, at a constant rate, stochastically changes its orientation or ``tumbles".
 In one dimension, and in presence of a drift, the equation of motion for an RTP reads
\bea
\dot{x}=v_{0}\,\sigma(t) + \mu \, .
\eea
 Here $v_0$ is the (constant) self-propulsion speed of the particle in the absence of the drift. The stochastic tumbling dynamics are encoded in
$\sigma(t)$, which is a telegraphic noise that switches between values $\pm 1$ at a (constant) rate $\gamma$. 
Let us recall that in the limit $\gamma \to \infty$ and $v_0 \to \infty$, keeping the ratio $v_0^2/  \gamma$ fixed, the RTP motion behaves like the one of a Brownian particle with a diffusion constant $D = v_0^2/ (2 \, \gamma)$.
Here  $\mu$ is the drift, and we assume that $|\mu| < v_0$, otherwise the particle always moves in the same direction and its local-time statistics behavior becomes rather trivial.

In the case of a RTP,  each time the particle crosses the origin, this contributes to the local time \eqref{localtime} precisely $1/|v|$ where $v$ is the particle's velocity at the time the origin is crossed. It is therefore easy to see that the local time can also be written as
\bea 
\mathcal{T} =   \frac{1}{v_0 + \mu} \, N_+(T) + \frac{1}{v_0 - \mu} \, N_-(T),
\eea
 where $N_{\pm}(T)$ are the number of crossings of the origin by the particle with positive and negative velocity, respectively,  in the time interval $[0,T]$. 
Denoting by $N(T) = N_+(T) + N_-(T)$ the total number of zero crossings, one clearly notes that if $N(T)$ is even then $N_{\pm}(T) = N(T)/2$ and one has
\be \label{localtime and crossing}
\mathcal{T} =\frac{v_0}{v_0^2 - \mu^2} \, N(T) \, .
 \ee
 In $N(T)$ is odd, Eq.~\eqref{localtime and crossing} is still approximately correct (in the limit $N \gg 1$). 
 From here it follows that one can study the distribution of $N(T)$ and deduce from it immediately the distribution of the local time $\mathcal{T}$.

We will show that the 
 local time follows an LDP in the large time limit, i.e  Eq. \eqref{LDP}.
 For this purpose we will compute the SCGF $\Lambda(k)$ for $N(T)$, 
 defined as
\bea \label{eqscgfN}
\Lambda(k) = \lim_{T \to \infty} \frac{1}{T} \,\, \ln \langle  e^{ k \, N(T)} \rangle \, .
\eea
 As we can see 
 from Eq.~\eqref{localtime and crossing}, it is related to the SCGF $\lambda(k)$ for $\mathcal{T}$, as
\be
\lambda(k) = \Lambda \Bigg (\frac{k \,  v_0}{v_0^2 - \mu^2} \Bigg ) \, .
\ee
Since $N(T)$ is bounded by the total number of tumbles $\mathcal{N}(T)$ (i.e $N(T) \leq \mathcal{N}(T)$),  it follows that $\left\langle e^{k N(T)}\right\rangle \le\left\langle e^{k\mathcal{N}(T)}\right\rangle$ (for $k \ge 0$).
As we now show, this gives a upper bound for $\Lambda(k)$.

The number of tumbles $\mathcal{N}$ follows a Poisson distribution  with mean $\gamma T$, whose moment generating function can be found explicitly
\be
\label{MGFPoisson}
\langle e^{k\,\mathcal{N}(T)}\rangle=\sum_{\mathcal{N}=0}^{\infty}\,\frac{1}{\mathcal{N}!}\,e^{k\,\mathcal{N}}\,e^{-\gamma\,T}\,\left(\gamma\,T\right)^{\mathcal{N}}=e^{\gamma\,T\,\left(e^{k}-1\right)}\,,
\ee
 which yields an upper bound for the SCGF of the number of zero crossings, 
\bea \label{upperbound}
\Lambda(k) \leq \gamma \, \Big (  e^{k}-1 \Big ) \, .
\eea
 More details and a lower bound are given in Appendix \ref{appendix:Ntumbles}.

 We will now show that  the probability $P_T(N)$ of  the number of crossings $N=N(T)$ follow a LDP at large $T$
 \be 
 \label{LDPNcrossings}
 P_T(N)  \sim \, e^{-T \, \tilde{I}(N / T)} ,
\ee
 and it can be connected to the local time using Eq. \eqref{localtime and crossing} as
 the LDP \eqref{LDP} holds and
  \bea
I\left(\mathcal{T}/T\right) =   \tilde{I}\left(\left(v_{0}^{2}-\mu^{2}\right)\mathcal{T}/v_{0}T\right) \, .
 \eea
  Using that $N(T) \le \mathcal{N}(T)$ together with the LDPs that both of these random variables satisfy, we show in Appendix \ref{appendix:Ntumbles}  that the rate function of $N(T)$ is bounded from  below by that of $\mathcal{N}(T)$
\be
\label{lowerboundI}
\tilde{I}\left(a\right)\ge a\ln\left(a/\gamma\right)-a+\gamma,
\ee
in the regime $a\ge \gamma$ (which corresponds to the right tail of the distribution of $\mathcal{N}(T)$).

We will now calculate the rate function $\tilde{I}$ and associated SCGF  $\Lambda$. To do so, we will consider the dynamics of the joint distribution of the position of the RTP and the number of zero crossings up to time $t$. The long-time behavior of the generating function of this joint distribution will yield  $\Lambda$, and then by applying the Legendre-Fenchel transform we will obtain $\tilde{I}$.

\begin{figure}
\centering
\includegraphics[width=0.9\linewidth]{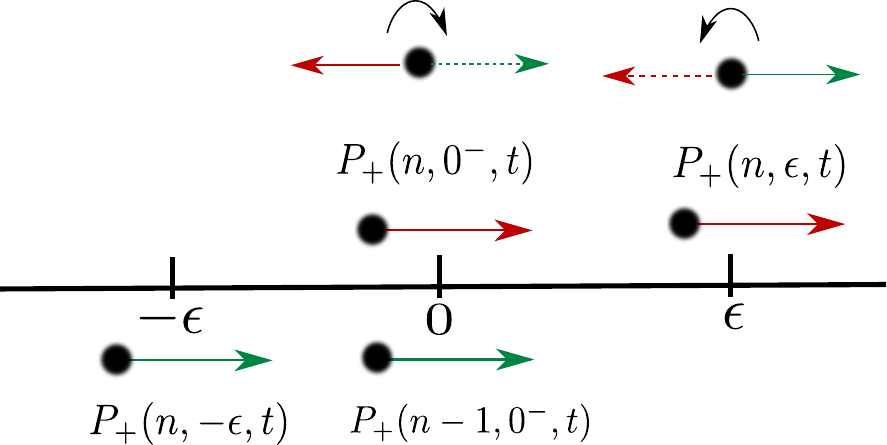}
\caption{
 Total flux of particles in the interval $[-\epsilon,\epsilon]$ with velocity $v_0 + \mu$. The six types of particles plotted correspond to the six terms on the right-hand side of Eq.~\eqref{fluxEpsilon}.}
\label{flux change}
\end{figure}

Let us define $P_\pm(n, x, t)dx$ to be the probability that at time $t$ the RTP is in the interval $[x, x + dx]$ with velocity $\pm v_0$ and has crossed the origin ($x = 0$)  exactly $n$ times. The Fokker-Planck equation that is satisfied by $P_\pm(n,x,t)$ in each of the two regions $x>0$ and $x<0$ is
\be \label{FP eq}
\partial_t \, \left(\begin{array}{c}
P_{+}\\
P_{-}
\end{array}\right) \, = \, L^{\dagger} \,  \left(\begin{array}{c}
P_{+}\\
P_{-}
\end{array}\right) ,
\ee
where $L^{\dagger}$ is the generator of the motion
\bea
L^{\dagger}=\left(\begin{array}{cc}
-(v_{0}+\mu)\partial_{x}-\gamma & \gamma\\[1mm]
\gamma & (v_{0}-\mu)\partial_{x}-\gamma
\end{array}\right) \, . \label{generator}
\eea
The dynamics of $n$ enter through the boundary conditions at $x=0$, which we now obtain.
The change in  time of the number of right-moving particles that have crossed the origin $n$ times and are in an interval $x \in [-\epsilon, \epsilon]$ is:
\bea
\label{fluxEpsilon}
&& \!\!\!\!\!\!\!\!\!\!\!\! \partial_t\, \int_{- \epsilon}^{\epsilon} P_+(n,x,t)  dx = (v_0+\mu) \Big [ P_+(n, -\epsilon, t)  \nn\\
&&  \quad + \, P_+(n-1, 0^-, t) -   P_+(n, 0^-, t) -  P_+(n, \epsilon, t) \Big ] \nonumber \\
&&  \quad + \, \gamma \int_{-\epsilon}^{\epsilon} \left[P_{-}\left(n, x, t\right) -P_{+}\left(n, x, t\right) \right]  dx \, .
\eea
 The six terms on the right-hand side of Eq.~\eqref{fluxEpsilon} correspond to
particles that enter or leave the interval $[-\epsilon,\epsilon]$, cross the origin, or tumble, see Fig. \ref{flux change}.
In the  limit $\epsilon \to 0$, the integrals in Eq.~\eqref{fluxEpsilon} vanish and  after replacing $\pm \epsilon$ by $0^\pm$ in the remaining terms we get the boundary condition for the right-moving particles,
\begin{eqnarray} \label{bc right}
     P_+(n-1, 0^-, t)  =  P_+(n, 0^+, t) \, .
\end{eqnarray}
Similarly, for  left-moving particles, one finds that the boundary condition  is
\bea \label{bc left}
 P_-(n-1, 0^+, t)  =  P_-(n, 0^-, t) \, .
\eea
 Note that $P_+(0, 0^+, t)=0$ and $P_-(0, 0^-, t)=0$. 

In order to find the SCGF $\Lambda$, it is useful to define the generating function of $P_\pm$ with respect to the number of crossings $n$
\be
\mathcal{G}_{k \pm} (x, t) = \sum_{n\geq 0} \, e^{n k}  P_\pm (n, x, t) \, .
\ee
At $x>0$ and at $x<0$, $\mathcal{G}_{k \pm} (x, t)$ satisfies a  Fokker-Planck equation with the same generator as in Eq. \eqref{FP eq}  and
Eq. \eqref{generator}
\be \label{generating fun}
\partial_t \, \left(\begin{array}{c}
\mathcal{G}_{k+}\\
\mathcal{G}_{k-}
\end{array}\right) \, = \, L^{\dagger}_k\, \left(\begin{array}{c}
\mathcal{G}_{k+}\\
\mathcal{G}_{k-}
\end{array}\right) \, ,
\ee
 with boundary conditions at $x=0$ that follow immediately from the boundary conditions  \eqref{bc right} and \eqref{bc left} above:
\begin{equation} \label{bc}
    e^k\,  \mathcal{G}_{k +}(0^-, t)  =  \mathcal{G}_{k +}( 0^+, t), \;\;  e^k\, \mathcal{G}_{k -} (0^+, t)  =  \mathcal{G}_{k -} ( 0^-, t)  \, .
\end{equation}
 Since these boundary conditions are $k$-dependent, we  have denoted the generator of the dynamics of $\mathcal{G}_{k \pm}$ by $L^{\dagger}_k$ (i.e., with a subscript $k$).
The solution of Eq. \eqref{generating fun} can be written by expanding $\mathcal{G}_{ k \pm}\left(x,t\right)$ in the basis of the eigenfunctions of the generator in Eq. \eqref{generator} 
\be
\label{Gksolexact}
\mathcal{G}_{k \pm}(x, t) = \sum_{i=0}^{\infty} \, c_i \, e^{\xi_i\, t} \,\, r_{k \pm}^{(i)} (x),
\ee
where $\xi_i$ and $ r_k^{(i)} (x)$ are the eigenvalues and  right eigenvectors of the generator respectively,  $L^{\dagger}_k r_k = \xi \, r_k$,
 with the boundary conditions \eqref{bc}. 
 Note that if the spectrum of $L^{\dagger}_k$ is continuous, then the sum \eqref{Gksolexact} becomes an integral.
 In the long-$T$ limit the term with largest eigenvalue $\xi_{\max}$ dominates in the above sum \eqref{Gksolexact}. 
Using that
\bea
&& \!\!\!\! \int_{-\infty}^{\infty}\left[\mathcal{G}_{k+}(x,t)+\mathcal{G}_{k-}(x,t)\right]dx \nonumber \\
&&\!\!\!\!  = \, \sum_{n\geq0}\,e^{nk}\,  \int_{-\infty}^{\infty}\, \left[P_{+}(n,x,t))+P_{-}(n,x,t)\right]dx
\eea
is the moment generating function  $\left\langle e^{kN(t)}\right\rangle $ of the number of zero crossings up to time $t$, we find that the SCGF is given by $\Lambda = \xi_{\max}$.
The problem thus boils down to finding the largest eigenvalue of the tilted operator  $L^\dagger_k$,
 where the tilt is in the boundary condition at $x=0$.

 It is convenient to study the  eigenvalue equations  by using the change of variables:
\begin{eqnarray}
    r_{k +}(x)+ r_{k -}(x)=\Phi_{k 1}(x)  \, ,\\ 
     r_{k +}(x) - r_{k -}(x)=\Phi_{k 2}(x) \, .
\end{eqnarray}
 The eigenvalue equations $L_{k}^{\dagger}\left(\begin{array}{c}
r_{k+}\\
r_{k-}
\end{array}\right)=\tilde{\Lambda}\left(\begin{array}{c}
r_{k+}\\
r_{k-}
\end{array}\right)$ then read, at $x>0$ and $x<0$,
\begin{eqnarray} \label{eigen value equations}
   \Tilde{\Lambda} \,  \Phi_{k 1}(x)&=&-v_{0} \,   \Phi'_{k 2}(x) -  \mu \,  \Phi'_{k 1}(x) \, , \\
     ( \Tilde{\Lambda}+2 \gamma )\,  \Phi_{k 2}(x)&=& -v_{0} \,  \Phi'_{k 1}(x) -  \mu \,   \Phi'_{k 2}(x) \, ,
     \label{eigen value equations right}
\end{eqnarray}
 and the boundary conditions \eqref{bc} become 
\begin{eqnarray} \label{bc left new}
    e^k\,  \Big ( \Phi_{k 1}(0^-) + \Phi_{k 2}(0^-) \Big ) &=&    \Phi_{k 1}(0^+) + \Phi_{k 2}(0^+) \, , \\   e^k\,  \Big ( \Phi_{k 1}(0^+) - \Phi_{k 2}(0^+) \Big ) &=&    \Phi_{k 1}(0^-) - \Phi_{k 2}(0^-) \, . \label{bc right new}
\end{eqnarray}
 In addition to the boundary conditions at $x=0$, the eigenfunctions  corresponding to the largest eigenvalue must also satisfy boundary conditions at $x \to \pm \infty$.
 Eqs. \eqref{eigen value equations} and \eqref{eigen value equations right} are linear, homogeneous, first-order ordinary differential equations with constant coefficients and may therefore be solved immediately in each of the two regimes $x>0$ and $x<0$. The solutions, that vanish at $x \to \pm\infty$, are given by
\bea
\label{Phik1Sol}
\Phi_{k 1}(x)&=&\begin{cases}
A_l\,  e^{\alpha_1 x} , & x < 0,\\[1mm]
A_r\,  e^{ \alpha_2 x}, & x > 0,
\end{cases} \\[1mm]
\Phi_{k 2}(x)&=&\begin{cases} - \frac{(\mu \alpha_1 + \Tilde{\Lambda})}{v_0  \alpha_1}
A_l\,  e^{\alpha_1 x} , & x < 0,\\[2mm]
- \frac{(\mu \alpha_2 + \Tilde{\Lambda})}{v_0  \alpha_2} A_r\,  e^{\alpha_2 x}, & x > 0,
\end{cases}
\eea
where 
\be
\alpha_{1,2}=\frac{1}{v_{0}^{2}-\mu^{2}}\,\left[\mu(\Tilde{\Lambda}+\gamma)\pm\sqrt{\gamma^{2}\,\mu^{2}+\Tilde{\Lambda}\,v_{0}^{2}\,(2\gamma+\Tilde{\Lambda})}\right]\,,
\ee
 where $\alpha_1$ ($+$ branch) is positive and $\alpha_2$ ($-$ branch) is negative.
The boundary conditions  \eqref{bc left new} and  \eqref{bc right new} above yield the following set of linear equations for the coefficients $A_l$ and $A_r$:
\begin{eqnarray}
   e^k\Bigg  (1 - \frac{\mu \alpha_1 + \Tilde{\Lambda}}{v_0  \alpha_1} \Bigg ) A_l  &=&  A_r \Bigg ( 1 - \frac{\mu \alpha_2 + \Tilde{\Lambda}}{v_0  \alpha_2} \Bigg ) \, ,\\[1mm]
    e^k \Bigg ( 1 + \frac{\mu \alpha_2 + \Tilde{\Lambda}}{v_0  \alpha_2}
 \Bigg ) A_r &=&  A_l \Bigg ( 1 + \frac{\mu \alpha_1 + \Tilde{\Lambda}}{v_0  \alpha_1}\Bigg ) \, .
\end{eqnarray}
 In order for these equations to have a nonzero solution, we must require them to be linearly dependent.  Working out this condition, we find that the
 combination $e^{2k}+ e^{-2k}+2=(2 \cosh k)^2$ massively simplifies, and obtain the following simple expression for the eigenvalue
\bea \label{scgf_mu}
\Tilde{\Lambda}{(k, \mu)} = \gamma \, \Bigg (  \frac{ \sqrt{v_0^2 - \mu^2}}{v_0} \, \cosh{k} - 1 \Bigg )  \, .
\eea
Similarly to the case of Brownian motion, here the SCGF  $\Lambda(k)$ cannot be negative. For zero drift,  the eigenvalue $\tilde{\Lambda}(k,\mu)$ in Eq.~\eqref{scgf_mu} is indeed nonnegative  so it gives the SCGF correctly, but for nonzero drift it may be negative. We therefore now analyze these two cases separately, starting from the zero-drift case. 
\begin{figure*}
\centering
\includegraphics[width=0.48\linewidth]{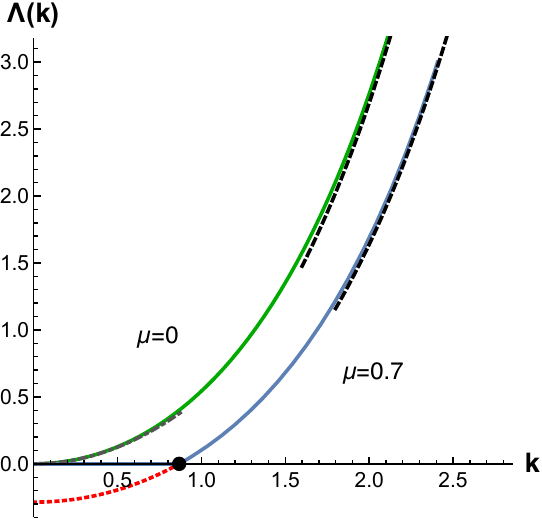}
\hspace{1mm}
\includegraphics[width=0.48\linewidth]{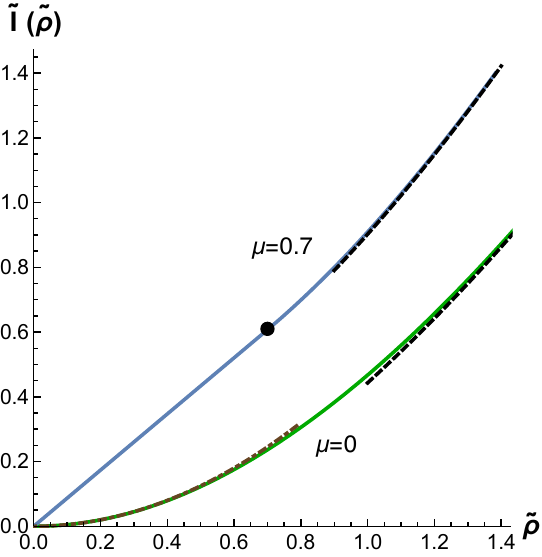}
    \caption{Plots of the SCGF and rate function for the number of times that the origin is crossed (which is proportional to the  local time) for an RTP  with $v_0 = \gamma = 1$ in the absence ($\mu = 0$) and presence  ($\mu = 0.7$) of drift. The black dashed lines denote the asymptotic behaviours  \eqref{LambdaLargek} and \eqref{ItildeLargeRho} which describe the limits $k \to \infty$ and $\rho \to \infty$ respectively. The red dashed line denotes the unphysical solution of the SCGF. The brown dot-dashed lines denote the 
    asymptotic behaviors $\Lambda(k \ll 1)$ and $\tilde{I}(\tilde{\rho} \ll 1)$ for the zero-drift case, see Eqs.~\eqref{asymptoticsLambda} and \eqref{asymptoticsItilde} respectively. 
    The solid circle denotes the critical point  which, for $\mu=0.7$, is at $k_c = 0.86730\dots$ corresponding to $\rho_c = 0.7$, see Eqs.~\eqref{kcSol} and \eqref{rhocSol} respectively.}
    \label{LDFswithdrift}
\end{figure*}

\subsection{No drift $\mu = 0$}

In the absence of any drift,  Eq.~\eqref{scgf_mu} gives the correct SCGF and it  simplifies to
\begin{equation}
\label{LambdaOfkSol}
   \Lambda(k) = \Tilde{\Lambda}(k, 0) = \gamma \,  (\cosh k - 1) \, .
\end{equation} 
The Legendre transform of this function yields the rate function for the number of zero crossings, 
\be
\label{IsolNcrossings}
\tilde{I}(\tilde{\rho}) = \gamma - \sqrt{\gamma^2+ \tilde{\rho}^2} +\tilde{\rho} \, \sinh^{-1}(\tilde{\rho}/\gamma)  \, .
\ee
 This rate function describes the large-deviation statistics of $N=N(T)$ through Eq.~\eqref{LDPNcrossings}.
 One can easily verify that $\Lambda(k)$ and $\tilde{I}(\tilde \rho)$ indeed satisfy the upper and lower bounds \eqref{upperbound} and \eqref{lowerboundI}, respectively.
The asymptotic behaviours of $\Lambda(k)$ and $\tilde I(\tilde{\rho})$ are given by
\bea
\label{asymptoticsLambda}
\!\!\! \Lambda(k) &\!\simeq& \! \begin{cases}
\!  \gamma \frac{k^2}{2} \,, & k \to 0\,,\\[2mm]
\!  \gamma ( \frac{e^k}{2} - 1 ) \,, & k \to \infty\,,
\end{cases}\\[2mm]
\label{asymptoticsItilde}
\!\!\! \tilde{I}(\tilde{\rho}) &\!\simeq& \! \begin{cases}
\!  \frac{\tilde{\rho}^{2}}{2 \gamma}-\frac{\tilde{\rho}^{4}}{24 \gamma}  \,, & \tilde{\rho}\to0\,,\\[2mm]
\! \tilde{\rho}\!\left[\ln\left(\frac{2\tilde{\rho}}{\gamma}\right)-1\right]+\gamma-\frac{\gamma^{2}}{4\tilde{\rho}}, & \tilde{\rho}\to\infty.
\end{cases}
\eea
 where $\tilde \rho \geq 0$. The function $\tilde I(\tilde \rho)$ reaches its minimum at $\tilde \rho=0$, which
corresponds to the typical and average number of crossings being sublinear in $T$. 
This SCGF and rate function  are shown in Fig. \ref{LDFswithdrift}. Both  functions are analytic,  i.e., they contain no singularities,  as in the case of the SCGF and rate function that describe the local-time distribution for a Brownian particles.
Moreover,
the rate function \eqref{IsolNcrossings} is in perfect agreement with the large-deviations asymptotic behavior that we extracted from the exact results of Ref.~\cite{SK21}, see Appendix \ref{appendix:Exact}.


\subsection{Nonzero drift $\mu \ne 0$}
For $\mu \neq 0$,  as in the case of the SCGF that describes the local-time statistics for Brownian motion,  $\tilde{\Lambda}(k)$ becomes negative below a critical value of $k=k_c$.  Following a similar argument to the one used above in Section \ref{BM with drift} and in \cite{NT2018, NT2018DPT}, the correct SCGF must be convex, and satisfy $\Lambda(0) = 0$ and $\Lambda'(0) = 0$. Since zero is also an eigenvalue of the tilted generator 
 (see Ref.~\cite{NT2018DPT} and Appendix \ref{appendix:LambdaZero}),
one finds that
$\Lambda\left(k\right)=\max\left\{ 0,\tilde{\Lambda}\left(k\right)\right\} $, i.e., (using Eq.~\eqref{scgf_mu}): 
\be 
\label{scgf_mu2}
\Lambda(k)=\begin{cases}
0\,, & k\leq k_{c}\,,\\[2mm]
\gamma\left(\frac{\sqrt{v_{0}^{2}-\mu^{2}}}{v_{0}}\,\cosh k-1\right)\,, & k>k_{c}
\end{cases}
\ee
 where 
\be
\label{kcSol}
k_{c}=\cosh^{-1}\left(\frac{v_{0}}{\sqrt{v_{0}^{2}-\mu^{2}}}\right)  = \coth^{-1}\left(\frac{v_{0}}{\mu}\right) \,.
\ee
The first derivative of SCGF thus shows a jump discontinuity and the system undergoes a first order dynamical phase transition at $k=k_c$. The subcritical phase corresponds to ``coexistence" (in time) between a ``phase" in which the particle is localized around the origin and a ``phase" in which it runs away to infinity due to the drift.

Using the Legendre-Fenchel transform \eqref{LegendreFenchel}, the rate function can be calculated:
\bea &&\!\!\!\! \tilde{I}(\tilde{\rho}) = \nn\\
&&
\!\!\!\!
\begin{cases}
\!\tilde{\rho}\,k_{c}, & \tilde{\rho}\leq\tilde{\rho}_{c},\\[2mm]
\!\gamma-\frac{\sqrt{v_{0}^{2}\left(\tilde{\rho}^{2}+\gamma^{2}\right)-\mu^{2}\gamma^{2}}}{v_{0}}+\tilde{\rho}\sinh^{-1} \! \left( \! \frac{\tilde{\rho}\,v_{0}}{\gamma\sqrt{v_{0}^{2}-\mu^{2}}} \! \right) \! , & \tilde{\rho}>\tilde{\rho}_{c}
\end{cases} \nonumber \\
\eea
where 
\be
\label{rhocSol}
\tilde{\rho}=\tilde{\rho}_c = \frac{\gamma \, \mu}{v_0},
\ee is the critical density that separates between the coexistence phase $0 < \tilde{\rho} < \tilde{\rho}_c$ and the localized phase $\tilde{\rho} > \tilde{\rho}_c$.
The rate function $\tilde{I}(\tilde{\rho})$ near $\tilde{\rho}_c$ behaves as 
\bea \label{eqratefunccritical}
&&\!\!\!\! \tilde{I}(\tilde{\rho})=\nn\\
&&\begin{cases}
\!\! \tilde{\rho}_{c}k_{c}+ k_{c}\,(\tilde{\rho}-\tilde{\rho}_{c}), & \tilde{\rho}<\tilde{\rho}_{c},\\[2mm]
\!\!\tilde{\rho}_{c}k_{c}+k_{c}(\tilde{\rho}-\tilde{\rho}_{c})+\frac{\left(\tilde \rho-\tilde \rho_{c}\right)^{2}}{2\gamma} + \dots, &  0 \leq  \tilde{\rho}-\tilde{\rho}_{c}\!\ll\!\tilde{\rho}_{c}.
\end{cases}\nn\\
\eea
hence it shows a jump discontinuity in the second derivative at $\tilde{\rho} = \tilde{\rho}_c$.

The nonzero drift $\mu$ makes it less likely for the RTP to perform many zero crossings (corresponding to $\tilde{\rho}=O(1)$). Therefore, as $\mu$ is increased, 
 $\tilde{I}(\tilde \rho)$ grows, as can indeed be seen in Fig.~\ref{LDFswithdrift}.

In the $k \to \infty$ and $\tilde{\rho} \to \infty$ limit, $\Lambda(k, \mu)$ and $\tilde{I}(\tilde{\rho})$ behaves as
\bea
\label{LambdaLargek}
\Lambda(k \to \infty) &\simeq& 
 \gamma \frac{\sqrt{v_0^2-\mu^2}}{v_0} \, \Bigg (\frac{e^k}{2} - 1 \Bigg)  \, , \\
\label{ItildeLargeRho}
\tilde{I}(\tilde{\rho} \to \infty) &\simeq&
\tilde{\rho}\left(\ln\frac{2v_{0}\tilde{\rho}}{ \gamma \sqrt{v_{0}^{2}-\mu^{2}}}-1\right)+\gamma-  \gamma^2 \frac{v_{0}^{2}-\mu^{2}}{4v_{0}^{2}\tilde{\rho}}\,.\nn\\
\eea
 It is easy to verify that $\Lambda(k)$ and $\tilde{I}(\tilde \rho)$ satisfy the upper and lower bounds given by Eqs.~\eqref{upperbound} and \eqref{lowerboundI} respectively.
 The SCGF and rate function for  $\mu=0.7$ are plotted in Fig.~\ref{LDFswithdrift}.

In the limit $\mu \to v_0^-$  the critical value $k_c$ behaves as
\be
k_{c}\simeq\frac{1}{2}\ln\left(\frac{2}{1-\frac{\mu}{v_{0}}}\right)\gg1,
\ee
and the SCGF  behaves as
 \be
\Lambda(k) \simeq \begin{cases}
0\,, & k\leq k_{c}\,,\\[2mm]
\gamma\left(e^{k-k_c} -1\right)\,, & k>k_{c} \, .
\end{cases}
\ee
Correspondingly, the critical value $\tilde{\rho}_c$ increases to $\tilde{\rho}_c \simeq \gamma$.
In this limit, any given value of $\tilde{\rho} = O(1)$ becomes very unlikely. In the subcritical regime, they are described by the linear part of the rate function
$\tilde{I} = k_c \tilde{\rho} \gg 1$. In the supercritical regime $\tilde{\rho} > \tilde{\rho}_c$, the fluctuations become  even more unlikely.

{\bf Remark}. For $\mu=0$ one can extract the large time $T$ behavior of the cumulants by taking  derivatives of \eqref{LambdaOfkSol} at $k=0$. One finds
that for even integer $q$ they behave as 
\be \label{cumT} 
\langle N(T)^q \rangle_c \simeq \gamma T \quad , \quad \langle {\cal T}^q \rangle =  \frac{\gamma T}{v_0^q}. 
\ee 
while the odd cumulants are sublinear in $T$ at large $T$. In presence of a drift $\mu>0$ since
 $\Lambda(k)$ vanishes identically for $k$ near $k=0$, all cumulants are sublinear in $T$.


\section{DV theory for RTPs} \label{DV for RTP}

 In this section, we develop a general DV theory for studying large deviations of dynamical observables in which the underlying stochastic process is the motion of an RTP 
 in the presence of an external potential. We will then apply this theory to study the distribution of the occupation time in a finite interval in the absence of drift, recovering  (in a non-trivial way) the local-time result given above in the limit of a very small interval.

\subsection{General}

Let us now consider a more general case in which the drift is not homogeneous in space:
\be
\dot{x}= \mu (x) + v_{0} \sigma(t) \, .
\ee
We aim to calculate the fluctuations of  dynamical observables \eqref{observable} 
in the large $T$ limit.
 Our starting point is
the generator $L^\dagger$ of the Fokker-Planck  (FP) equation
$   \left(\begin{array}{c}
\dot{P}_{+}\\
\dot{P}_{-}
\end{array}\right)  =  L^{\dagger}  \left(\begin{array}{c}
P_{+}\\
P_{-}
\end{array}\right) 
$
for the joint  probability density function (PDF) of the position and orientation of the particle. This generator is the same $2\times2$ matrix as the one given above in
\eqref{generator}, except that the drift now may be position dependent: 
 \bea
L^{\dagger}=\left(\begin{array}{cc}
-\partial_{x} (v_{o}+\mu(x))-\gamma & \gamma\\[1mm]
\gamma & \partial_{x} (v_{o}-\mu(x)) -\gamma
\end{array}\right) \, .
\eea

Now we write the FP equation for the joint PDF, $P_{\pm}(x,A,t)$, of the particle's position, orientation, and of 
$A(t) = \int_0^t f(x(t'))dt'$: 
\be
\left(\begin{array}{c}
\dot{P_{+}}\\
\dot{P_{-}}
\end{array}\right) = L^\dagger \left(\begin{array}{c}
P_{+}\\
P_{-}
\end{array}\right) 
-\left(\begin{array}{cc}
f(x) & 0\\
0 & f(x)
\end{array}\right)\left(\begin{array}{c}
\partial_{A}P_{+}\\
\partial_{A}P_{-}
\end{array}\right) \, .
\ee
 In order to deal with the derivatives wrt $A$, it is useful to  define the following double sided Laplace transform
\be
G_{k \pm} (x,t)= \int_{- \infty}^{\infty} e^{k A} \, P_\pm(x, A, t) \, dA \, ,
\ee
 whenever the integral converges.
 The 
 time evolution of the functions $G_{k \pm}$ is then given by
\be
\left(\begin{array}{c}
\dot{G}_{k+}\\
\dot{G}_{k-}
\end{array}\right)=  \mathcal{L}_k^\dagger \left(\begin{array}{c}
G_{k+}\\
G_{k-}
\end{array}\right),
\ee
where
\be
\mathcal{L}_k^\dagger = L^\dagger + k\left(\begin{array}{cc}
f(x) & 0\\
0 & f(x)
\end{array}\right) \, .
\ee
In the long time limit,  the solution to this equation behaves as
\be \label{asympt} 
G_{k \pm} (x,t) \sim e^{\lambda t} \,   \Psi_{k \pm} (x),
\ee
where $\lambda$ and $\Psi_k  =\left(\begin{array}{c}
\Psi_{k+}\\
\Psi_{k-}
\end{array}\right)$ are the dominant eigenvalue and eigenfunction of  $\mathcal{L}_k^\dagger$ respectively
(which depend on $k$, although for convenience we will make their $k$ dependence implicit). 
The temporal dependence in this last equation determines the SCGF $\lambda(k)$ of the distribution of the observable 
$\mathcal{A}$ (Note that $\mathcal{A} = A(t=T)$.)

 Unlike the case of Brownian motion in an external potential, it is not possible to symmetrize the operator $\mathcal{L}_k^\dagger$. From a physical point of view, this reflects the fact that the RTP dynamics are not symmetric under time reversal \cite{Hugo2018}.

 The eigenvalue problem,
 $\mathcal{L}_k^\dagger\Psi_k(x)=\lambda(k)\,\Psi_k(x)$, may be slightly simplified by applying the change of variables
 \bea
\Psi_{k+}(x)+\Psi_{k-}(x)&=&\Phi_{1}(x), \\
\Psi_{k+}(x)-\Psi_{k-}(x)&=&\Phi_{2}(x) .
\eea
In these variables the eigenvalue problem takes the form
\be
\label{QPhiLambda}
\mathcal{Q}\left(\begin{array}{c}
\Phi_{1}\\
\Phi_{2}
\end{array}\right)=\lambda\left(\begin{array}{c}
\Phi_{1}\\
\Phi_{2}
\end{array}\right),
\ee
where 
 \be
\mathcal{Q}=\left(\begin{array}{cc}
-\partial_{x} \mu\left(x\right)+kf(x) & -v_{o}\partial_{x}\\[1mm]
-v_{o}\partial_{x} & -\partial_{x} \mu\left(x\right) +kf(x)-2\gamma
\end{array}\right) \, .   
\ee

 Moreover, if $f(x) = f(-x)$ is mirror symmetric, and if there is no drift, $\mu(x)=0$, the operator $\mathcal{Q}$
commutes with the operator 
$\left(\Phi_{1}(x),\Phi_{2}(x)\right)\to\left(\Phi_{1}(-x),-\Phi_{2}(-x)\right)$ which corresponds to an inversion of space.
Hence $(\Phi_1(x), \Phi_2(x))$ is an eigenvector of the latter operator. It follows that one of the $\Phi_i$'s is symmetric  $\Phi_i(x) = \Phi_i(-x)$
and the other is antisymmetric
 $\Phi_i(x) = -\Phi_i(-x)$.

\begin{figure*}
\centering
\includegraphics[width=0.45\linewidth]{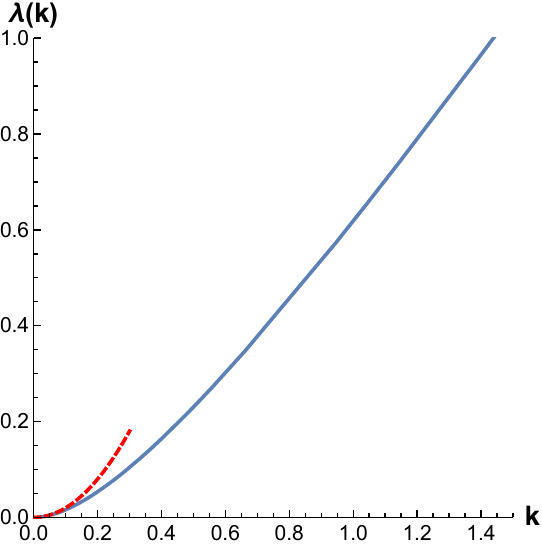}
\hspace{1mm}
\includegraphics[width=0.45\linewidth]{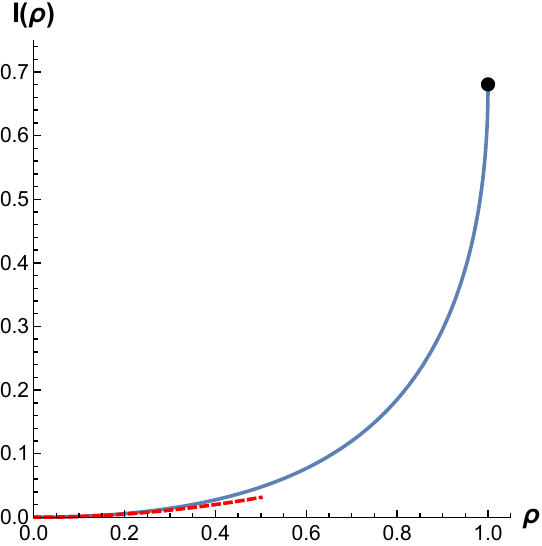}
\caption{ Plots of the SCGF and rate function for occupation time of a RTP in  the interval $a=1$ for $\gamma=v_0=1$.  Solid lines correspond to the exact solution, for which $\lambda(k)$ is obtained from the solution to the transcendental equation \eqref{LambdaOfkOccupation}, and $I(\rho)$ is given by the Legendre transform of $\lambda(k)$. Dashed lines correspond to the asymptotic behaviors \eqref{lambdaSmallk} and \eqref{ISmallrho}. The solid circle denotes the  value $I(1) = 0.680977\dots$   
which corresponds to the survival rate of the RTP inside the interval $[-1,1]$ (see Appendices \ref{I(1) calculation} and \ref{survival prob} for more details). }
\label{fig_occupationtime} 
\end{figure*}

 We now apply this method to the particular cases of local time and occupation time.
 For the particular case $f(x) = \delta(x)$ (corresponding to the local time)  a direct application of this method encounters a technical difficulty, because  the product $\mathcal{Q}\left(\begin{array}{c}
\Phi_{1}\\
\Phi_{2}
\end{array}\right)$ contains products of the delta function $\delta(x)$ with functions that are not continuous at $x=0$,  and such products are not well defined.
We will therefore demonstrate our general method now by solving the more general problem of the  distribution of the occupation time in an interval of length $2a$. The results for the local time can then be recovered by carefully taking the limit $a\to 0$, and as we will show below, they are found in perfect agreement with those of the previous section. For simplicity, we will work out the case of no drift $\mu=0$.
 This case is technically simpler because one can exploit the mirror symmetry properties described above.
For nonzero drift, one can in principle solve in a similar manner. The solution is quite cumbersome, but qualitatively it is expected to exhibit a DPT similar to the ones described above for the local time of an RTP and the local or occupation time of a Brownian particle \cite{NT2018, NT2018DPT}.

\subsection{Occupation time}

 We now apply the general DV method that we formulated above, to the particular case of the occupation time of the RTP in the interval $[-a,a]$, corresponding to Eq.~\eqref{occupation},  i.e. to the choice $f(x)=\mathds{1}_{[-a,a]}(x) $.

 In each of the two regions $x\in\left[-a,a\right]$ and $x\notin\left[-a,a\right]$, the eigenvalue problem \eqref{QPhiLambda} is a set of two first-order linear ordinary differential equations with constant coefficients. It is therefore straightforward to obtain the general solution. Since $f(x)=f(-x)$ is mirror symmetric, following the discussion above, it is sufficient to consider solutions for which exactly one of the two functions $\Phi_1(x),\Phi_2(x)$ is symmetric, and the other is anti-symmetric.

It turns out that the solution to \eqref{QPhiLambda} corresponding to the dominant eigenvalue $\lambda(k)$
is such that $\Phi_1(x) = \Phi_1(-x)$ is symmetric and
$\Phi_2(x) = -\Phi_2(-x)$ is anti-symmetric \cite{footnote:conditional}.
The general solution that satisfies these properties and decays at $x\to \pm \infty$ is 
\bea
\label{Phi1}
\Phi_{1}(x)&=&\begin{cases}
  E\, \cos{\alpha x} \, , & -a\leq x\leq a\, ,\\[1mm]
C \, e^{\beta x}\, , & -a\leq x\, ,\\[1mm]
C \, e^{-\beta x}\, , & x\geq a\, ,
\end{cases}\\[1mm]
\Phi_{2}(x)&=&\begin{cases}
\frac{v_{0}\alpha E}{\lambda(k)-k+2\gamma} \, \sin{\alpha x}\, , & -a\leq x\leq a\, ,\\[2mm]
\frac{v_{0}\beta C}{-\lambda(k)-2\gamma} \, e^{\beta x}\, , & -a\leq x\, ,\\[2mm]
 \frac{v_{0}\beta C}{\lambda(k)+2\gamma} \, e^{-\beta x}\, , & x\geq a\, ,
\end{cases}
\eea
where 
\bea
\alpha^{2}&=&\frac{(2\gamma-k+\lambda(k))(k-\lambda(k))}{v_{0}^{2}} \, ,\\
\beta^{2}&=&\frac{(\lambda(k)+2\gamma)\lambda(k)}{v_{0}^{2}} \, .
\eea
Note that $\alpha^2$ may be negative, and in that case, inside the interval $[-a,a]$, $\Phi_1$ and $\Phi_2$ are given by hyperbolic cosine and  hyperbolic sine, respectively, as functions of $x$.

The $\Phi_i$'s should   be continuous at the boundaries $x=\pm a$ (note however that their derivatives $d\Phi_i/dx$ are not continuous at $x=\pm a$), which gives  a set of linear homogeneous equations for the coefficients $E$ and $C$
\bea
E\cos\alpha a &=& C  e^{-\beta a}  \, ,\\
\frac{v_{0}\alpha E}{\lambda(k)-k+2\gamma}\sin\alpha a&=&  \frac{v_{0}\beta C}{\lambda(k)+2\gamma}e^{-\beta a}  \, .
\eea

For a nontrivial solution to exist, we get the following conditions:  
\bea
\label{LambdaOfkOccupation}
&& \!\!\!\!\!\!\!\!\!\!\!\!\!\!\!\!\!\!\!\!\!\!\!\! \tan^{2}\sqrt{\left[2\gamma-k+\lambda(k)\right]\left[k-\lambda(k)\right]}\frac{a}{v_{0}} \nn\\
&&\qquad\qquad\qquad =\frac{\lambda(k)\left[\lambda(k)-k+2\gamma\right]}{\left[\lambda(k)+2\gamma\right]\left[k-\lambda(k)\right]},
\eea
 if $0<k-\lambda(k)<2\gamma$,
and
\bea
\label{LambdaOfkOccupation2}
&& \!\!\!\!\!\!\!\!\!\!\!\!\!\!\!\!\!\!\!\!\!\!\!\! \tanh^{2}\sqrt{\left[k-\lambda\left(k\right)-2\gamma\right]\left[k-\lambda(k)\right]}\frac{a}{v_{0}} \nn\\
&&\qquad\qquad\qquad =\frac{\lambda(k)\left[k-\lambda\left(k\right)-2\gamma\right]}{\left[\lambda(k)+2\gamma\right]\left[k-\lambda(k)\right]},
\eea
if $k-\lambda(k) > 2\gamma$.
 Eq.~\eqref{LambdaOfkOccupation2} is only needed for sufficiently small $a$'s, because at large enough intervals, $k-\lambda(k)$ cannot exceed the value $2 \gamma$.
 The solution to these transcendental equations yields the SCGF $\lambda(k)$.  The equations can be solved numerically, or, alternatively, one can obtain $\lambda(k)$ in a parametric form, see  Apppendix \ref{SCGFofOccupation} for the details. Using the Legendre-Fenchel transformation of  $\lambda(k)$, we get the rate function $I(\rho)$.  We plot the SCGF and rate function in Fig.~\ref{fig_occupationtime}. As evident from the figures of the SCGF and the rate function, there are no singularities and the functions are analytic functions, i.e., no DPT occurs.

 The leading behaviors of $\lambda(k)$ and $I(\rho)$ at small $k$ read
\bea \label{lambdaSmallk}
&& \lambda(k)= 2 \frac{a^2 \gamma}{v_0^2} k^2 +O(k^3),   
\\
&& I(\rho) = \frac{ v_0^2 \rho^2}{8 \gamma a^2}   + O(\rho^3). \label{ISmallrho}
\eea 
A few higher order terms, as well as up to the fourth cumulant of  occupation time ${\cal R}$ \eqref{occupation} are given in the Appendix \ref{SCGFofOccupation}. For  $\rho \to 1 $ (corresponding to large $ k \to \infty$) the behaviour of the rate function depends on $a$. For details  see Appendix \ref{I(1) calculation}. This $I(\rho = 1)$ can also be interpreted as the long-time survival rate of the RTP inside the interval $[-a,a]$ (see Appendix \ref{survival prob}). 
It is given by the solution to the transcendental equation
\be  
\label{I1MT}
\frac{a \gamma}{v_0} = \frac{\cos^{-1}(-1 + I(1)) }{2 \sqrt{I(1) (2-I(1))}}.
\ee
for $\tilde{a} > 1/2$,  where $\tilde a= a \gamma/v_0$ (see  analogous formula for $\tilde{a} < 1/2$ in Appendix \ref{I(1) calculation}).

Note that the rate functions depend on  the single dimensionless parameter $\tilde a$. We now discuss separately
the two limits $\tilde a \gg 1$ and $\tilde a \ll 1$

\subsection{Large $a$ limit: Brownian limit} 

 In the limit $a \gg v_0 / \gamma$, the particle typically experiences many tumbling events before travelling a distance that is of order of the size of the interval. Therefore, one expects that in this limit our result \eqref{LambdaOfkOccupation} should recover the existing one for occupation time of a Brownian particle in an interval.

This is indeed the case. Physically, the simplest way to see it is by considering Eq.~\eqref{LambdaOfkOccupation} in the limit $v_{0}\rightarrow\infty$ and $\gamma\rightarrow\infty$
keeping the ratio $\frac{v_{0}^{2}}{2\gamma}=D$ to be fixed (note that in this limit one indeed has $a \gg v_0 / \lambda$). In this limit, $k$ and $\lambda(k)$ are negligible compared to $\gamma$, and Eq.~\eqref{LambdaOfkOccupation} simplifies to
\be
\tan^{2}\sqrt{\frac{k-\lambda(k)}{D}}\;a=\frac{\lambda(k)}{k-\lambda(k)} \, ,
\ee
 which indeed coincides with the corresponding result in Ref.~\cite{NT2018DPT}.

\subsection{Small $a$ limit: Local time limit} 

 In the opposite limit, $a \ll v_0 / \gamma$, it becomes extremely unlikely for the particle to tumble while it is in the interval $[-a,a]$. Assuming that no such tumbling events occur, each time the origin is crossed, the amount of time spent by the particle in the interval $[-a,a]$ is (exactly) $2a/v_0$, so we find that the occupation time is approximately $\mathcal{R}\simeq 2a N/v_0$. To remind the reader, $N$ is the number of times that the origin is crossed and it in turn is proportional to the local time through $\mathcal{T} = N/v_0$. We therefore expect the SCGFs of the occupation time and number of crossings to be related via
\be
\lambda\left(\frac{k'v_{0}}{2a}\right)\simeq\Lambda\left(k'\right) \, .
\ee

Let us now show that in the limit $a \ll v_0 / \gamma$, our result \eqref{LambdaOfkOccupation2} recovers the local-time result derived in the previous section. Replacing $k=\frac{k'v_{0}}{2a}$  in Eq.~\eqref{LambdaOfkOccupation2} and then taking the limit $a \to 0$  while $\lambda,\gamma$ remain of order one, the equation simplifies to
\be
\label{smallaLimittanh2}
\tanh^{2}\frac{k'}{2}\simeq\frac{\lambda}{\lambda+2\gamma} \, .
\ee
It is easy to verify that the solution to Eq.~\eqref{smallaLimittanh2} is indeed given by $\lambda = \Lambda(k')$ as given in Eq.~\eqref{LambdaOfkSol}.

\subsection{Distribution of the endpoint conditioned on $\rho$}

Physically, $\Phi_1(x)$ (properly normalized) gives the distribution of the RTP's position at time $T$, conditioned on a given value of $\mathcal{R}$, cf. \cite{CH15}. From Eq.~\eqref{Phi1} one finds that if $\alpha^2 >0$ then this distribution is 
unimodal  and
maximal at the origin, whereas if $\alpha^2 <0$ it is bimodal  and
maximal at the
edges of the interval.
 As we will see below, this change of behavior occurs only when $\tilde a = a \gamma/v_0 < 1/2$ and
in that case for $\rho=\rho_c(\tilde a)$ given by
\be 
\rho_c(\tilde a) = \frac{8 \tilde a^2 \left(3-4  \tilde a^2\right)}{16 \tilde a^4+3},  \quad 
\ee 
so that for $1 \geq \rho>\rho_c(\tilde a)$ the distribution is bimodal, while for $\rho < \rho_c(\tilde a)$
it is unimodal (note that at $\rho=\rho_c(\tilde a)$ the distribution is uniform in the interval). 
We stress that this change of behavior does not correspond to a phase transition as $\lambda(k)$ and $I(\rho)$
are analytic around this point.

\section{Summary and Discussion} \label{conclusion}

In this paper, we have investigated the fluctuations of generic dynamical variables in which the underlying stochastic process is an RTP, with or without an external drift, in one dimension.  We worked out in detail  two important particular examples: The local time spent by the RTP at a given spatial point, and the occupation time  of an interval by a RTP. We have shown that the fluctuations  of these observables satisfy  a 
LDP in the long time limit, and calculated the associated rate functions exactly.

For the  fluctuations of the local time, we achieved this result by using a simple relation between the local time and the number of times that the RTP crosses the origin. Interestingly, we observed that the rate function associated with the local time fluctuations is nonanalytic in the presence of an external drift. We interpret this singularity as a dynamical phase transition of first order. 
A very similar situation occurs for drifted Brownian motion, both for the local time, and for the occupation time \cite{NT2018}.
The transition occurs between two regimes (or ``phases"): a localized regime, in which the particle \sout{is} remains in the vicinity of the origin throughout the entire dynamics, and a coexistence regime, in which the particle stays near the origin for a fraction of the dynamics and then escapes to infinity due to the drift.

In order to study the occupation time statistics, we extended the DV method, which was previously formulated for Brownian motion \cite{Hugo2018}, to the case of RTPs. This method gives a theoretical framework for studying fluctuations of general dynamical observables of the type \eqref{observable}, by mapping the problem to that of finding the largest eigenvalue of a linear operator. The eigenfunctions give information on the PDF of the final position of the particle conditioned on a given value of the observable \eqref{observable}. The rate function $I(\rho =1)$ gives the survival rate of the RTP inside the interval $[-a,a]$. 
 For the case $\mathcal{R}=T$ corresponding to survival in the interval $[-a,a]$ we observed that, in the absence of an external drift, the eigenfunctions change behaviour at $\tilde{a} = 1/2$, from a unimodal at $\tilde{a} > 1/2$ to bimodal at $\tilde{a} < 1/2$. 
 For other values of $\mathcal{R}$ a similar phenomenon occurs, but the change of behavior is at a different value of $\tilde{a}$.
Finally, we recovered the local time results by carefully taking the limit $a \to 0$.

Several future directions of research remain.
It would be interesting to further understand the large-deviations behavior for the local time and occupation time for RTP by characterizing the conditioned process, i.e., the dynamics of the RTP conditioned on a given value of local time or occupation time,  in greater detail than in the present work \cite{MM22, MBE19}. 
It would also be interesting to study other dynamical observables of the type \eqref{observable}, for RTPs that are perhaps trapped in external potentials. 
One could also study dynamical observables that are  of a more general form than the type \eqref{observable} \cite{Hugo2018} or even study more detailed characterizations of the process such as the empirical position distribution (so-called level 2 large deviations \cite{HNE16, CVC22, Monthus24}).
The DV formalism can be extended for other types of active particles, and perhaps in higher dimension (e.g., active Brownian particles \cite{BLLRVV2016}).
It would also be interesting to study multi-particle  systems, with possible interactions between the particles \cite{LMS19, LMS21, Singh21, ARYL21, PBDN21, RSBI22, Cates22, ARYL22, KPS23, MukherjeeSmith23, SmithMeerson24}.

\subsection*{Acknowledgments}

NRS acknowledges support from the Israel Science Foundation (ISF) through Grant No. 2651/23.
 PLD acknowledges support from 
ANR grant ANR-23-CE30-0020-01 EDIPS,
and thanks KITP for hospitality, 
supported by NSF Grants No. NSF PHY-1748958 and PHY-2309135.

\appendix

\section{Relation between number of zero crossings and number of tumbles} 
\label{appendix:Ntumbles}

As mentioned in the main text, the number of tumbles $\mathcal{N}$ follows a Poisson distribution with mean $\gamma T$. Using Eq.~\eqref{MGFPoisson}, one finds that the SCGF of $\mathcal{N}$ is given by 
\be
\lambda_{\text{tumbles}}\left(k\right)=\gamma\left(e^{k}-1\right)\,.
\ee
It then follows that $\mathcal{N}$ satisfies an LDP with rate function that is obtained by applying the Legendre-Fenchel transform to this SCGF, and is given by
\be
I_{\text{tumbles}}\left(a\right)=a\ln\left(a/\gamma\right)-a+\gamma \, .
\ee

Since the number of zero crossings $N$ satisfies $N \le \mathcal{N}$, the SCGF and rate function of $\mathcal{N}$ yield upper and lower bounds (respectively) for those of $N$, as we now show.
From $N \le \mathcal{N}$ it follows that, for any $k>0$ and any given realization of the RTP, $e^{kN}\le e^{k\mathcal{N}}$. Therefore also
$\left\langle e^{kN}\right\rangle \le\left\langle e^{k\mathcal{N}}\right\rangle $, from which it follows that
$\Lambda\left(k\right)\le\lambda_{\text{tumbles}}\left(k\right)$
which is the upper bound given in Eq.~\eqref{upperbound} of the main text.
To obtain a lower bound for the rate function of $N$, we note that $N \le \mathcal{N}$ implies that
\be
\label{lowerboundNm}
\text{Prob}\left(N=m\right)\le\text{Prob}\left(\mathcal{N}\ge m\right) \, .
\ee
Now, using that $N$ and $\mathcal{N}$ both satisfy LDPs, with rate functions $\tilde{I}\left(a\right)$ and $I_{\text{tumbles}}\left(a\right)$ respectively, we can approximate the left- and right-hand sides of Eq.~\eqref{lowerboundNm} by
$e^{-T \tilde{I}\left(m/T\right)}$
and 
$\int_{m}^{\infty}e^{-T\tilde{I}\left(m'/T\right)}dm'$
respectively.
In the large-$T$ limit, the integral over $m'$ is in the leading order equal to the maximum of the integrand over all values of $m'$. It then follows that
\be
\label{lowerboundIwithmin}
\tilde{I}\left(a\right)\ge\min_{a'\ge a}I_{\text{tumbles}}\left(a'\right) \, .
\ee
The global minimum of the rate function of $\mathcal{N}$ is 
$I_{\text{tumbles}}\left(a=\gamma\right)=0$. Hence, for $a<\gamma$, Eq.~\eqref{lowerboundIwithmin} gives the trivial lower bound
$\tilde{I}\left(a\right)\ge 0$.
However, since $I_{\text{tumbles}}\left(a\right)$ is monotonically increasing at $a > \gamma$, it follows that for $a>\gamma$ Eq.~\eqref{lowerboundIwithmin} simplifies to give the lower bound
$\tilde{I}\left(a\right)\ge I_{\text{tumbles}}\left(a\right)$
which is Eq.~\eqref{lowerboundI} of the main text.

\smallskip

\section{Calculation of SCGF for the occupation time statistics} \label{SCGFofOccupation}

Here we solve the transcendental equations Eq. \eqref{LambdaOfkOccupation} and \eqref{LambdaOfkOccupation2} for the SCGF of $\mathcal{R}$ in a parametric form. 
 The parametric solution will be given as the union of three branches.
Let us begin from the former (first two branches). We define
\be
\label{Bdef}
B=(2\,\gamma-k+\lambda) \, (k-\lambda) .
\ee
We will obtain $\lambda(k)$ in a parametric form, such that $\lambda$ and $k$ will both be given explicitly as functions of $B$.
Solving Eq.~\eqref{Bdef} for $k-\lambda$, we obtain
\be
\label{kMinusLambdaofB}
 k-\lambda(k)=\gamma\:\pm\sqrt{\gamma^{2}-B} \, .
\ee
$k-\lambda$ is not a single-valued function of $B$. (Real) solutions to Eq.~\eqref{kMinusLambdaofB} exist for $0\leq B\leq\gamma^{2}$. At $B=0,$ solutions
are $0$ and $2\gamma$  and as $B \to \gamma^{2}$, both the solutions become equal to $\gamma$.
The range $0< k-\lambda < \gamma$, is described by the first solution:
\bea
\label{kMinusLambdaofB1}
k-\lambda(k)=\gamma-\sqrt{\gamma^{2}-B}.
\eea
Plugging Eq.~\eqref{kMinusLambdaofB1} into \eqref{LambdaOfkOccupation} we get
\be
\tan^{2}\sqrt{B}\frac{a}{v_{0}}=\frac{\lambda(k)\left(\gamma+\sqrt{\gamma^{2}-B}\right)}{\left[\lambda(k)+2\gamma\right]\left(\gamma-\sqrt{\gamma^{2}-B}\right)} \, ,
\ee
and solving this equation for $\lambda$ we obtain  the first branch
\be
\label{LambdaofB1}
\lambda=\frac{2\gamma\left(\gamma-\sqrt{\gamma^{2}-B}\right)\sin^{2}\left(\frac{\sqrt{B}a}{v_{0}}\right)}{\sqrt{\gamma^{2}-B}+\gamma\cos\left(\frac{2\sqrt{B}a}{v_{0}}\right)},
 \ee
 where $B$ varies in $B \in [0,\gamma^{2}]$ and 
$k$ is then obtained explicitly as a function of $B$ by the two equations \eqref{kMinusLambdaofB1} and \eqref{LambdaofB1}.

Similarly, the range $\gamma< k-\lambda < 2\gamma$ is described by the second solution:
\bea
k-\lambda(k)=\gamma+\sqrt{\gamma^{2}-B},
\eea
which, following similar steps as in the previous case, yields  the second branch
\bea \label{LambdaofB2}
\lambda=\frac{2\gamma\left(\sqrt{\gamma^{2}-B}+\gamma\right)\sin^{2}\left(\frac{\sqrt{B}a}{v_{0}}\right)}{\gamma\cos\left(\frac{2\sqrt{B}a}{v_{0}}\right)-\sqrt{\gamma^{2}-B}} \,,
\eea
 where $B$ varies in $B \in [0,\gamma^{2}]$ (it decreases back from $\gamma^2$ to $0$).

The range $k - \lambda > 2\gamma$ is described by the solution to Eq.~\eqref{LambdaOfkOccupation2}. 
 Now $B$ is negative and we write
\be
\label{Bdef2}
|B|=- B = (k-\lambda-2\,\gamma)\,(k-\lambda) \, .
\ee
Solving again for $k-\lambda$, we obtain
\bea
k-\lambda(k)=\gamma+\sqrt{\gamma^{2}+|B|},
\eea
where we required $k-\lambda(k) > 0$.
Plugging the last equation into \eqref{LambdaOfkOccupation2} and solving for $\lambda$, we obtain  the third branch
\bea \label{LambdaofB3}
\lambda=\frac{2 \, \gamma\left(\sqrt{|B|+ \gamma^2}+\gamma\right)\sinh^{2}\left(\frac{\sqrt{|B|}a}{v_0}\right)}{\sqrt{|B|+\gamma^2}-\gamma \, \cosh\left(\frac{2\sqrt{|B|}a}{v_0}\right)} \,,
\eea
 where $|B|$ increases from $0$ to $+\infty$.
As in the previous cases, $k$ is given as a function of $B$ by combining the last two equations.

 Now that we have identified the three branches we must examine when they are relevant. This is
discussed in terms of the dimensionless parameter 
\be 
\tilde a = \frac{a \gamma}{v_0}.
\ee 
Upon plotting $k(B)$ and $\lambda(B)$ (not shown) we find that 
(i) for $0< \tilde a <1/2$
all three branches are relevant but the third branch is relevant only for $0< |B|/\gamma^2 < |b_c^{(3)}(\tilde a)|$
given in the next subsection in \eqref{root3}, 
since at this value of $B$ the denominator in \eqref{LambdaofB3} vanishes,
(ii) for $1/2< \tilde a <\pi/4$
only the first two branches are relevant but the second branch is relevant only for $1> B/\gamma^2 > b_c^{(2)}(\tilde a)$
given in \eqref{root2}, 
since at this value of $B$ the denominator in \eqref{LambdaofB2} vanishes (iii) for  $\tilde a >\pi/4$
only the first branch is relevant but only for $0< B/\gamma^2 < b_c^{(1)}(\tilde a)$
given in \eqref{root1} since at this value the denominator in \eqref{LambdaofB1} vanishes.

The vanishing of these denominators shows that at these special values of $B/\gamma^2$ 
both $k=k(B)$ and $\lambda=\lambda(B)$ diverge. 
As analyzed in detail in the next subsection Sec. \ref{I(1) calculation}, it corresponds to the point $\rho=1$ and determines the value of $I(1)$.
Hence one can state alternatively that the possible range of $k-\lambda$ depends on $I(1)$ as $0 < k-\lambda< I(1)$ (see Sec. \ref{I(1) calculation} for more details). Thus if $I(1) < \gamma$ then only the first solution is relevant, if $\gamma < I(1) < 2\gamma$ then both first solutions are relevant and if $I(1) > 2\gamma$ then all three solutions play a role.
$I(1)$ describes the long-time survival probability of the RTP in the interval $[-a,a]$, and therefore the value of $I(1)$ decreases as a function of $a$.

 The Legendre-Fenchel transform reduces to the Legendre transform because $\lambda(k)$ is convex and differentiable. Using this the rate function $I(\rho)$ is also obtained in a parametric form by the formulas:
\bea
\rho(B) &=& \frac{d\lambda}{dk} = \frac{d\lambda / dB}{dk /db}  = \frac{d\lambda / dB}{dk /dB} = \frac{\lambda'(B)}{k'(B)} \, , \\
I(B) &=& \rho(B) k(B) - \lambda(B) \, .
\eea
These formulas are of course applied separately 
 to each of the three branches for $\lambda(B)$ and $k(B)$ 
described above, and the result is
a smooth (analytic) rate function $I(\rho)$.
In Fig.~\ref{fig_occupationtime} of the main text we plot $\lambda$ vs. $k$ and the corresponding rate function $I(\rho)$. 

 These formula allow to obtain the small argument expansion of $\lambda(k)$
and $I(\rho)$ to high orders, which leads to the cumulants of ${\cal R}$.
For simplicity here we express $\lambda$, $k$ and $I$ in units of $\gamma$ 
(since they are inverse times) while we note that $\rho={\cal R}/T$ is dimensionless.
These dimensionless quantities then depend on the dimensionless
parameter (i.e. the reduced interval size) 
\be
\tilde a= \frac{a \gamma}{v_0}.
\ee 
All above formula still apply, as it amounts to set $\gamma=1$. The
 small $k$ expansion gives (using the first branch) 
\be \label{lambdaexp2} 
 \lambda( k)= 
2 \tilde{a}^2 {k}^2-\frac{16
   \tilde{a}^4
   {k}^3}{3}+\frac{2}{45}
   \tilde{a}^4 \left(368
   \tilde{a}^2+15\right)
   {k}^4+O({k}^5),
\ee
from which the four lower cumulants of ${\cal R}$ can be obtained, 
and  
\be
I(\rho) =  
\frac{{\rho}^2}{8
   \tilde{a}^2}+\frac{{\rho}^3}
   {12 \tilde{a}^2}+\frac{\left(352
   \tilde{a}^2-15\right)
   {\rho}^4}{5760
   \tilde{a}^4}+O\left({\rho}^5
   \right) \, .
\ee

Taking derivatives of \eqref{lambdaexp2} w.r.t. $k$ and restoring units, we obtain the large time behavior of the four lowest cumulants of
the occupation time as 
\bea 
&& \langle {\cal R}^2 \rangle_c = 4 \gamma \tilde{a}^2 T + o(T) \, , \\
&& \langle {\cal R}^3 \rangle_c = -32 \gamma \tilde{a}^4 T + o(T) \, ,\\
&& \langle {\cal R}^4 \rangle_c = \gamma \tilde{a}^4 (16 + \frac{5888}{15} \tilde{a}^2)   T + o(T) \, .
\eea  
 Note that the leading behavior at small $\tilde a$ is compatible with ${\cal R} \simeq 2 a {\cal T}$
and the values given in \eqref{cumT} for the cumulants of ${\cal T}$ (the odd cumulants being subdominant).
\\

 Finally note that the analysis of the branches performed above allows to locate the
position of the change of behavior from unimodal to bimodal endpoint distribution
discussed in the text. Indeed, it occurs at the intersection point of the second and 
third branch as $B \to 0$, and thus it occurs only for $\tilde a<1/2$ since 
for $\tilde a>1/2$ the third branch is no longer relevant and the distribution
is always unimodal. This leads to 
$\rho= \lambda'(B)/k'(B)|_{B=0} = \rho_c(\tilde a) :
= \frac{8 \tilde a^2 \left(3-4  \tilde a^2\right)}{16 \tilde a^4+3}$ as
indicated in the main text. Equivalently one can say that
the endpoint distribution is bimodal for $\tilde a < \tilde a_c(\rho)$
given by
\be 
\tilde a_c(\rho) = \sqrt{ \frac{3 - \sqrt{3} \sqrt{3 -\rho
   ^2-2 \rho }}{4 \rho
   +8} },
   \ee 
   where $\tilde a_c(\rho)$ varies from $0$ to $1/2$ as $\rho$ increases from $0$ to $1$.

\subsection{ Calculation of $I(\rho=1)$ \label{I(1) calculation}}

Since the occupation time cannot be larger than the total time, the ratio $\rho= {\cal R}/T$
lies in $[0,1]$. The limit $\rho \to 1$ corresponds to $k, \lambda(k) \to +\infty$ but such that
the difference $k-\lambda(k)$ reaches a finite limit. This in turns means that $B$ also reaches a finite asymptotic value $B \to B_c$. This 
asymptotic value $B_c$ is a function of the dimensionless parameter $\tilde a = a \gamma/v_0$. 
Depending on whether $k-\lambda(k)$ converges to
a value in $[0,\gamma]$, $[\gamma,2 \gamma]$ or $[2 \gamma,+\infty]$, one must use the
first, second and third branches respectively. These three cases
corresponds to three different ranges of values for $\tilde a$. Everywhere below
we denote $B=\gamma^2 b$, and use the dimensionless quantities (which amounts
to set $\gamma=1$). 

Consider the case of the first branch. Then 
\be 
\lim_{k \to +\infty} k - \lambda(k) = 1 - \sqrt{1- b},
\ee 
where $b \in [0,1]$, 
and we recall that 
\be
\label{LambdaofB1new}
\lambda= \lambda(b) = \frac{2 \left(1-\sqrt{1-b}\right)\sin^{2}( \tilde a \sqrt{b} )}{\sqrt{1-b}+ \cos(2 \tilde a \sqrt{b})}.
 \ee 
One sees that as $b$ increases $\lambda(b)$ increases and diverges
for $b=b_c$ such that the denominator in the above expression for $\lambda(b)$ vanishes.
Hence  $b_c=b_c^{(1)}(\tilde a)$ is the smallest positive root of 
\be 
\sqrt{1-b_c} + \cos(2 \tilde a \sqrt{b_c}) = 0  \label{root1}.
\ee 
One can invert this equation to obtain 
$\tilde a=\tilde a(b_c)$ as a function of $b_c$ 
\be
\tilde a = \frac{1}{2 \sqrt{b_c}} \arccos( - \sqrt{1-b_c} ). 
\ee 
As $b_c$ increases from $0$ to $1$, one sees that $\tilde a$ decreases from $+\infty$ (
with $\tilde a \simeq \pi/(2 \sqrt{b_c})$ for small $b_c$) to 
$\tilde a_1=\frac{\pi}{4}$. The first branch is thus relevant to study $\rho \to 1$ when $\tilde a \in [\tilde a_1,+\infty[$. 
Next, using the relation between $\tilde a$ and $b_c$
the pole of $\lambda(b)$ at $b=b_c$ become
\be 
\lambda(b)  \simeq \frac{g(b_c)}{b_c-b}, 
\ee
 where $g(b_c)$
is positive for $b_c \in [0,1]$. On the other hand one can rewrite, for $b \in [0,b_c]$ 
\bea  
&& \frac{k(b)}{k'(b)} \lambda'(b) - \lambda(b) 
\\
&& = (\lambda(b) +1 - \sqrt{1-b}) \frac{\lambda'(b)}{\lambda'(b) + \frac{1}{2 \sqrt{1-b}} }-\lambda(b) \nn \\
&& \simeq 1 - \sqrt{1-b_c} - \frac{b_c-b}{2 \sqrt{1- b_c} }, \nn 
\eea 
using that $\lambda(b)/\lambda'(b) \simeq (b_c-b)$
From it one obtains
\bea  \label{I1expression}
&& I(1) = \lim_{k \to + \infty} ( k \lambda'(k) - \lambda(k) ) = \lim_{b \to b_c}
\frac{k(b)}{k'(b)} \lambda'(b) - \lambda(b) \nn \\
&& =  1 - \sqrt{1-b_c}. 
\eea  
Once again one can invert the relation between $I(1)\in [0,1]$ and $\tilde a \in [\frac{\pi}{4},+\infty]$ and obtain
\be  \label{I1-1}
\tilde a = \frac{\cos^{-1}(-1 + I(1)) }{2 \sqrt{I(1) (2-I(1))}},
\ee 
which is Eq.~\eqref{I1MT} of the main text.
The asymptotic behavior at large $\tilde a$ corresponds to the limit $I(1) \to 0$. In this limit, Eq.~\eqref{I1-1} simplifies to $\tilde{a}\simeq\frac{\pi}{\sqrt{8I(1)}}-\frac{1}{2}$. Inverting this relation we obtain (we now restore the units)
\be \label{I1 large a}
I(1)= \frac{ \pi^2}{8 \tilde a^2} \left( 1 - \frac{1}{\tilde a}+ O(\frac{1}{\tilde a^2}) \right) = \frac{\pi^2 D}{4 a^2 } \left(1 - \frac{v_0^2}{\gamma a} + O(\frac{1}{a^{2}}) \right),
\ee 
where $D=v_0^2/(2 \gamma)$. The leading order agrees with the decay rate 
of the Brownian particle with absorbing walls at $\pm a$ and diffusion
coefficient $D$ (indeed the ground state eigenfunction in that case is 
$\cos( \frac{\pi x}{2a} )$ which applying $D \partial_x^2 $ 
gives the decay rate $\frac{\pi^2 D}{4 a^2}$).

Let consider now the second branch. 
\be
\lambda=\frac{2 (\sqrt{1-b}+1) \sin^{2}(\tilde a \sqrt{b} )}
{\cos(2 \tilde a \sqrt{b})-\sqrt{1-b} }  \quad , \quad 
k-\lambda=1+\sqrt{1-b},
\ee
 where $b$ varies $b \in [0,1]$ from $1$ down to $0$. 
The asymptotic value  $b=b_c=b_c^{(2)}(\tilde a)$ is now determined by 
\be
\cos(2 \tilde a \sqrt{b_c})= \sqrt{1-b_c}, \quad  \quad \tilde a = \frac{1}{2 \sqrt{b_c}} \arccos(  \sqrt{1-b_c} ). \label{root2} 
\ee 
This corresponds to the interval $\tilde a \in [\frac{1}{2}, \frac{\pi}{4}]$, where $b_c=0$ corresponds to $\tilde a=1/2$.
One finds now
\be 
\lambda(b)  \simeq \frac{g(b_c)}{b-b_c}, 
\ee
where  $g(b_c)$
is positive for $b \in [0,1]$. This leads again to
\be 
I(1) = 1 + \sqrt{1- b_c} + O(b-b_c), 
\ee 
which using the relation between $b_c$ and $\tilde a$ leads to the
relation for $\tilde a \in [\frac{1}{2}, \frac{\pi}{4}]$ 
\be \label{I1 2}
\tilde a = \frac{\cos^{-1}(-1 + I(1)) }{2 \sqrt{I(1) (2-I(1))}}.
\ee 
This relation is identical to the one from branch 1 except that now
$I(1)$ varies between $1$ and $2$. For $I(1)=2$ one has $\tilde a=1/2$.

\begin{figure}
     \centering
\includegraphics[scale=0.6]{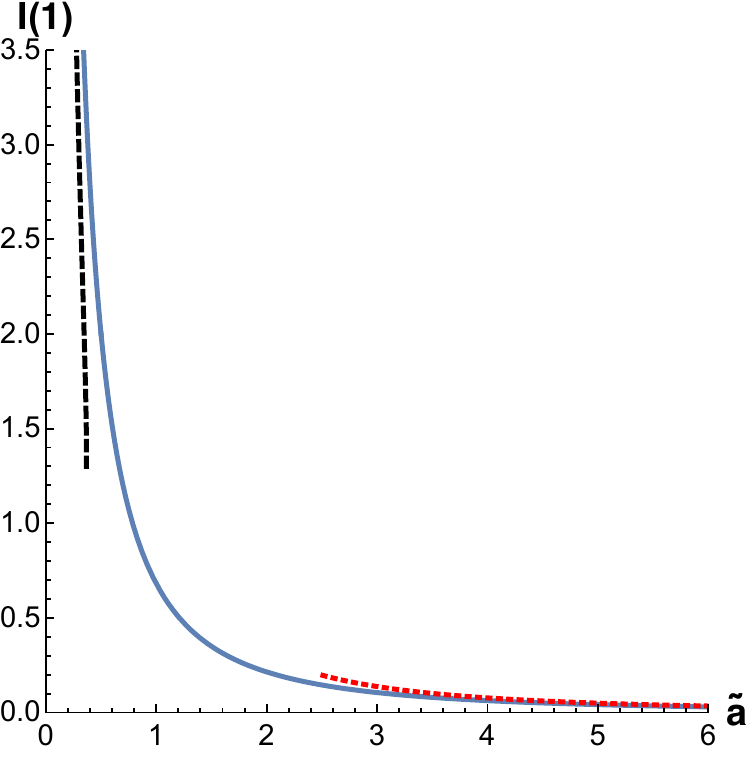}
\caption{ $I(\rho=1)$ as a function of $\tilde{a}$ given by Eqs. \eqref{I1 2} and \eqref{I1-3}. The black dashed and red dotted line denotes the asymptotic behaviours of $I(1)$ at $\tilde{a} \to 0$ (Eq. \eqref{I1 small a}) and $\tilde{a} \to \infty$ (the leading-order approximation $I(1)\simeq\frac{\pi^{2}}{8\tilde{a}^{2}}$ from Eq. \eqref{I1 large a}) limit respectively. }
\label{fig:I1}
 \end{figure}

Let us consider now the third branch where $b$ is negative. One has
\be
\lambda=\frac{2 (\sqrt{1+|b|}+1) \sinh^{2}(\tilde a \sqrt{|b|} )}
{\sqrt{1+|b|} - \cosh(2 \tilde a \sqrt{|b|}) },  \quad  \quad 
k-\lambda=1+\sqrt{1+|b|},
\ee
 where $b<0$ varies $|b| \in [0,+\infty]$. 
The asymptotic value  $b=b_c=b_c^{(3)}(\tilde a)$ is now determined by 
\be
\cosh(2 \tilde a \sqrt{|b_c|})= \sqrt{1+|b_c|}, \quad  \quad \tilde a = \frac{\cosh^{-1}(  \sqrt{1+|b_c|} ) }{2 \sqrt{|b_c|}}. \label{root3}
\ee 
This corresponds to the interval $\tilde a \in [0,\frac{1}{2}]$, where $|b_c|=+\infty$ corresponds to $\tilde a=0$.
One finds now
\be 
\lambda(b)  \simeq \frac{g(|b_c|)}{|b_c|-|b|}, 
\ee
where $g(|b_c|)$
is positive for $b \in [0,+\infty]$. This leads again to
\be 
I(1) = 1 + \sqrt{1 +  |b_c|} + O(b-b_c), 
\ee 
which using the relation between $b_c$ and $\tilde a$ leads to the
relation for $\tilde a \in [0,\frac{1}{2}]$ 
\be  \label{I1-3}
\tilde a = \frac{\cosh^{-1}(-1 + I(1)) }{2 \sqrt{I(1) (I(1)-2}},
\ee 
where now
$I(1)$ varies between $2$ and $+\infty$. For $I(1)=2$ one has $\tilde a=1/2$.
The asymptotics for large $I(1)$ is
\be 
\tilde a = \frac{\ln(2 I(1))}{2 I(1)} (1 + \frac{1}{I(1)} - \frac{1}{I(1) \ln(2 I(1))} + \dots), 
\ee 
leading as $\tilde a \to 0$ to the asymptotics
\be \label{I1 small a}
I(1) \simeq \frac{\ln(\frac{1}{\tilde a})}{2 \tilde a} \, .
\ee 
 $I(1)$ is plotted, as a function of $\tilde{a}$, in Fig.~\ref{fig:I1}, together with its asymptotic behaviors.

\section{Survival probability in an interval} \label{survival prob}

 In this appendix, we calculate the long-time survival probability of an RTP in an interval. This probability decays in time with rate $I(1)$, which we now calculate in a different method to the one presented in Appendix \ref{I(1) calculation}.

Let us denote  $S_\sigma(x,t)$ the probability that a RTP initially in the interval $[-a,a]$ remains
inside this interval for all times up to time $t$, and is at position $x$ in state  $\sigma\in\left\{ +,-\right\} $ at time $t$.
At large $t$ one expects that $S_\pm(x,t) \sim S_\pm(x) e^{- r t}$. Here $r>0$ is the survival rate $r = - \lim_{T \to +\infty} \ln S(T)$, which
does not 
depend on the details of the initial condition. It is given by the smallest eigenvalue of the linear
problem
\be 
- r S_{\sigma}(x) = - v_0 \sigma \partial_x S_{\sigma}(x)  + \gamma S_{-\sigma}(x) - \gamma S_{\sigma}(x),
\ee 
with $\sigma=\pm 1$, with boundary conditions $S_{+}(-a)=S_{-}(a)=0$. One looks for
a solution such that $S(x)=S_+(x)+S_-(x)$ is even in $x$ and $R(x)=S_+(x)-S_-(x)$ is odd in $x$.
The boundary condition becomes simply $S(a)=R(a)$. Let us again  use the dimensionless
units (equivalent to setting $\gamma=v_0=1$).
One finds that for $r <2 $ and $x \in [-a,a]$
\be S(x) = A \cos (\alpha x), \quad  \quad R(x) = B \sin (\alpha x),
\ee 
with $A/B= \alpha/r = (2 -r)/\alpha$, leading to $\alpha^2 = r (2  - r)$ and to
the equation which determines $r$
\be 
\tan^{2}\left(\tilde{a}\sqrt{r(2-r)}\right)=\frac{2-r}{r},
\ee 
which upon inversion leads to
\be 
\tilde a = \frac{\cos^{-1} (-1 + r)}{2 \sqrt{r(2-r) }},
\ee 
where $r$ decreases from $r=2$ to $r=0$ as $\tilde a$ increases with $\tilde a \in [1/2,+\infty]$. 
One sees from \eqref{I1-1} that it coincides with the first two
branches for $I(1)$,
i.e. $I(1)=r$. The third branch of $I(1)$ is obtained by considering $r>2$
in which case 
\be S(x) = A \cosh (\alpha x) \quad , \quad R(x) = B \sinh (\alpha x),
\ee 
with $A/B= \alpha/r = (r -2)/\alpha$, leading to $\alpha^2 = r (r  - 2)$ and to
the equation which determines $r$
\be 
\tanh^{2}\left(\tilde{a}\sqrt{r(r-2)}\right)=\frac{r-2}{r},
\ee 
which upon inversion leads to
\be 
\tilde a = \frac{\cosh^{-1} (-1 + r)}{2 \sqrt{r(r-2) }},
\ee 
where $r$ decreases from $r=+\infty$ to $r=2$ as $\tilde a$ increases with $\tilde a \in [0,1/2]$.
We see that again $r=I(1)$ from \eqref{I1-3}. 

The function $S(x)$, upon normalization in the interval $[-a,a]$
then gives the probability to find the particle at position $x$  at time $t$, conditioned upon 
its long time survival  up to time $t$. We see that it takes quite different shapes depending
on the value of $\tilde a$ with a change of behavior at $\tilde a=1/2$.
For small intervals the particle will be found closer to the boundaries (quite different from the Brownian), while 
for large intervals it will be found far from the boundaries (as for the Brownian).

\medskip

\section{Comparison of the local time statistics with the exact result of Ref. \cite{SK21}}
\label{appendix:Exact}

 In Ref. \cite{SK21}, an exact result was obtained for the local time of a run and tumble particle in the absence of drift, i.e. for $\mu=0$.
It was found that the probability $P_T(N)$ that the particle crosses the origin $N$ times up to time $T$ is given by 
\be \label{PTN} 
P_{T}\left(N\right)=e^{-\gamma T}\left[I_{N}\left(\gamma T\right)+I_{N+1}\left(\gamma T\right)\right] \, ,
\ee
where $N \geq 0$ and $I_N$ is the modified Bessel function. This formula is exact for all $T$ and for
a particle starting at the origin \cite{footnote:N}.

We are interested in the limiting behavior of $P_T(N)$ in \eqref{PTN} for 
$T \to \infty$ with $N/T = \tilde{\rho}$ held constant.
We use the asymptotic formula \cite{BesselAsymptoticsDLMF}
\be
I_{\nu}\left(\nu z\right)\sim e^{\nu\eta},\quad\eta=\sqrt{1+z^{2}}+\ln\frac{z}{1+\sqrt{1+z^{2}}} \, ,
\ee
 which holds for fixed $z$ in the limit $\nu \to \infty$,
we find that
\bea
&& \!\! P_T\left(N\right) \sim \nn\\
&&\!\! \exp\left[-T\left(\gamma-\tilde{\rho}\left(\sqrt{1+\left(\frac{\gamma}{\tilde{\rho}}\right)^{2}}+\ln\frac{\left(\frac{\gamma}{\tilde{\rho}}\right)}{1+\sqrt{1+\left(\frac{\gamma}{\tilde{\rho}}\right)^{2}}}\right)\right)\right] \nn\\
&&\!\! = \exp\left[-T\left(\gamma-\sqrt{\tilde{\rho}^2+\gamma^2}+ \tilde{\rho} \, \sinh^{-1}{\frac{\tilde{\rho}}{\gamma}}\right)\right],
\eea
 which is precisely the LDP \eqref{LDPNcrossings} where the rate function coincides with $\tilde{I}(\tilde \rho)$ in Eq.~\eqref{IsolNcrossings}.

 The derivation presented in this appendix is shorter than the derivation of Eq.~\eqref{IsolNcrossings} in the main text. However, the latter derivation presented is straightforward to extend to the case of nonzero drift $\mu \ne 0$, and more generally, the methods that we use in the main text are expected to be applicable in a relatively broad range of scenarios, since they do not rely on exact results at finite $T$.

\section{Eigenfunctions for the eigenvalue $\Lambda=0$ in the local time problem}
\label{appendix:LambdaZero}

Here we show that, at all $k \ge 0$, $\Lambda=0$ is an eigenvalue of the tilted generator $L_k^\dagger$ that corresponds to the problem of local time that an RTP spends in the vicinity of the origin.
In the spirit of Ref.~\cite{NT2018DPT} in which a very similar situation arises involving drifted Brownian motion, we look for a solution to Eqs.~\eqref{eigen value equations} and \eqref{eigen value equations right} of the form
\bea
\label{Phik1Lambda0}
\Phi_{k1}(x)&=&\begin{cases}
A\,e^{\alpha x}+B, & x<0,\\[1mm]
1, & x>0,
\end{cases}\\[1mm]
\Phi_{k2}(x)&=&\begin{cases}
Ce^{\alpha x}, & x<0,\\[1mm]
0, & x>0
\end{cases}
\eea
and with zero eigenvalue. Eqs.~\eqref{eigen value equations} and \eqref{eigen value equations right} are trivially satisfied at $x>0$, and from the regime $x<0$, they yield the relations
\bea
\label{ABC1}
C&=&-\mu A/v_{0}, \\
C&=&-\alpha v_{0}A/\left(2\gamma+\alpha\mu\right),
\eea
respectively.
Comparing the latter two conditions, we obtain
$\alpha=2\mu\gamma/\left(v_{0}^{2}-\mu^{2}\right)$.
The boundary conditions \eqref{bc left new} and \eqref{bc right new} at $x=0$ become
\bea
\label{ABC2}
A+B+C&=&e^{-k}\,,\\
\label{ABC3}
A+B-C&=&e^{k}\,.
\eea
Eqs.~\eqref{ABC1}, \eqref{ABC2} and \eqref{ABC3} are a set of linear equations for the coefficients $A,B,C$.

The solution to these equations is
\bea
A&=&\frac{v_{0}}{\mu}\sinh k,\\
B&=&\cosh k-\frac{v_{0}}{\mu}\sinh k,\\
C&=&-\sinh k\, .
\eea
We note that $A>0$ at all $k > 0$, and $B>0$ at $0 < k < k_c$ where $k_c$ is the criticial point given in Eq.~\eqref{kcSol} of the main text. We thus find that in the subcritical regime $0 < k < k_c$, $\Phi_{k1}$ in Eq.~\eqref{Phik1Lambda0} is positive. In the supercritical the solution Eq.~\eqref{Phik1Lambda0} ceases to be positive for all $x$, but the physical solution in this regime is the localized eigenfunction given by Eq.~\eqref{Phik1Sol} of the main text.


The solution given in this appendix, with eigenvalue $\Lambda=0$, exists at all $k\ge0$, and so does the solution given in the main text, with the eigenvalue given in Eq.~\eqref{scgf_mu}. For nonzero drift, as explained in the main text, the true SCGF $\Lambda(k)$ is given by Eq.~\eqref{scgf_mu2}, which is the \emph{largest} eigenvalue of the tilted operator.


\begin{thebibliography} {99}

\bibitem{WF1973} G. Wilemski and M. Fixman, \textit{General theory of diffusion‐controlled reactions},  {\href{https://pubs.aip.org/aip/jcp/article/58/9/4009/764812/General-theory-of-diffusion-controlled-reactions} {J. Chem. Phys. 58, 4009
(1973)}.}
\bibitem{Doi1975} M. Doi, \textit{Theory of diffusion-controlled reaction between non-simple molecules. I}, {\href{https://www.sciencedirect.com/science/article/pii/0301010475800437} {Chem. Phys., 11, 107 (1975)}}.
\bibitem{BCKMO2005} O. Benichou, M. Coppey, J. Klafter, M. Moreau, and G.
Oshanin, \textit{Mean joint residence time of two Brownian particles in a sphere}, {\href{https://iopscience.iop.org/article/10.1088/0305-4470/38/33/001} {J. Phys. A: Math. Gen. 38, 7205 (2005)}}.

\bibitem{Koshland1980}D. E. Koshland, \textit{Bacterial Chemotaxis as a Model Behavioral System} (Raven, New York, 1980).

\bibitem{KK2021} G. Kishore and A. Kundu, \textit{Local time of an Ornstein–Uhlenbeck particle},  {\href{https://iopscience.iop.org/article/10.1088/1742-5468/abe93d} {J. Stat. Mech.  033218 (2021).}}
\bibitem{CTB2010}  S. Carmi, L. Turgeman and E. Barkai,  \textit{On Distributions of Functionals of Anomalous Diffusion Paths}, {\href{https://doi.org/10.1007/s10955-010-0086-6}{J Stat Phys 141, 1071–1092 (2010).}} 
\bibitem{SHB2009} F. D. Stefani, J. P. Hoogenboom, and E. Barkai, \textit{Beyond Quantum Jumps: Blinking Nano-scale Light Emitters}, {\href{https://pubs.aip.org/physicstoday/article/62/2/34/399157/Beyond-quantum-jumps-Blinking-nanoscale-light} {Physics Today 62 nu. 2, p. 34 (February 2009).}}
\bibitem{MC2002} S. N. Majumdar and A. Comtet, \textit{Local and Occupation Time of a Particle Diffusing in a Random Medium}, {\href{https://journals.aps.org/prl/abstract/10.1103/PhysRevLett.89.060601} { Phys. Rev. Lett. 89, 060601 (2002).}}
\bibitem{SMC2006}S. Sabhapandit, S. N. Majumdar, and A. Comtet, \textit{Statistical properties of functionals of the paths of a particle diffusing in a one-dimensional random potential}, 
{\href{https://journals.aps.org/pre/abstract/10.1103/PhysRevE.73.051102} {Phys. Rev. E 73, 051102 (2006)}}.
\bibitem{CDM2002} A. Comtet, J. Desbois, and S. N. Majumdar, \textit{The local time distribution of a particle diffusing
on a graph}, {\href{https://iopscience.iop.org/article/10.1088/0305-4470/35/47/102}{J. Phys. A: Math. Gen. 35 L687 (2002)}}.
\bibitem{PCRK2019} A. Pal, R. Chatterjee, S. Reuveni, and A. Kundu, \textit{Local time of diffusion with stochastic resetting}, {\href{https://iopscience.iop.org/article/10.1088/1751-8121/ab2069} {J. Phys. A: Math. Theor. \textbf{52} 264002 (2019)}}.
\bibitem{BMR23} I. N. Burenev, S. N. Majumdar, and A. Rosso, \textit{Local time of a system of Brownian particles on the line with steplike initial condition},  {\href{https://journals.aps.org/pre/abstract/10.1103/PhysRevE.108.064113} {Phys. Rev. E \textbf{108}, 064113 (2023)}.}
\bibitem{SmithMeerson24} N. R. Smith, B. Meerson, \textit{Macroscopic fluctuation theory of local time in lattice gases}, {\href{https://www.sciencedirect.com/science/article/pii/S0378437124001249?via%3Dihub} {Physica A \textbf{639}, 129616 (2024)}}.



\bibitem{Schweitzer2003} F. Schweitzer, \textit{Brownian Agents and Active Particles:
Collective Dynamics in the Natural and Social Sciences},
{\href{https://link.springer.com/book/10.1007/978-3-540-73845-9} {Springer: Complexity, Berlin, (2003)}}.
\bibitem{RBELS2012}  P. Romanczuk, M. Bar, W. Ebeling, B. Lindner, and
L. Schimansky-Geier, \textit{Active Brownian particles}, {\href{https://link.springer.com/article/10.1140/epjst/e2012-01529-y}{Eur.
Phys. J. Special Topics 202, 1 (2012)}}
\bibitem{MJRLPRS2013}M. C. Marchetti, J. F. Joanny, S. Ramaswamy, T. B.
Liverpool, J. Prost, M. Rao, and R. Aditi Simha, \textit{Hydrodynamics of soft active matter}, {\href{https://journals.aps.org/rmp/abstract/10.1103/RevModPhys.85.1143}{Rev. Mod. Phys. 85,
1143 (2013).}}
\bibitem{FGGVWW2015} \'{E}. Fodor, M. Guo, N. Gov, P. Visco, D. Weitz, and F.
van Wijland, \textit{Activity-driven fluctuations in living cells},
{\href{https://iopscience.iop.org/article/10.1209/0295-5075/110/48005}{Europhys. Lett. 110, 48005 (2015)}}.
\bibitem{GKKRST23} O. Granek, Y. Kafri, M. Kardar, S. Ro, A. Solon, J. Tailleur, \textit{Inclusions, Boundaries and Disorder in Scalar Active Matter}, {\href{
https://doi.org/10.48550/arXiv.2310.00079
}{arXiv:2310.00079}.}



\bibitem{Santiago2018} I. Santiago, \textit{Nanoscale active matter matters: Challenges and opportunities for self-propelled nanomotors}, {\href{https://doi.org/10.1016/j.nantod.2018.01.001}{Nano Today, \textbf{19}, 11 (2018)}}.
\bibitem{GXGG2020} A. Ghosh, W. Xu, N. Gupta and D. H. Gracias, \textit{Active matter therapeutics}, {\href{https://doi.org/10.1016/j.nantod.2019.100836}{Nano Today, \textbf{31}, 100836 (2020)}}

\bibitem{SK21} P. Singh and A. Kundu, \textit{Local time for run and tumble particle}, {\href{https://journals.aps.org/pre/abstract/10.1103/PhysRevE.103.042119}{Phys. Rev. E \textbf{103}, 042119 (2021)}}.

\bibitem{DZ1998} A. Dembo, O. Zeitouni. \textit{Large Deviations Techniques and Applications}. {\href{http://www.springer.com/us/book/9783642033100}{Springer, New York, 2nd
edition (1998)}.}
\bibitem{Hollander2000} F. den Hollander. \textit{Large Deviations. Fields Institute Monograph}, {\href{http://bookstore.ams.org/fim-14.s}{AMS, Providence (2000)}} 


\bibitem{Hugo2009} H. Touchette, \textit{The large deviation approach to statistical mechanics}, {\href{https://doi.org/10.1016/j.physrep.2009.05.002} {Phys. Rep. \textbf{478}, 3 (2009).}}
\bibitem{Hugo2018} H. Touchette,  \textit{Introduction to dynamical large deviations of Markov processes.} {\href{https://www.sciencedirect.com/science/article/pii/S0378437117310567?via%3Dihub} {Physica A: Statistical Mechanics and its Applications \textbf{504}, 5-19 (2018).}}

\bibitem{NT2018} P. Tsobgni Nyawo, and H. Touchette. \textit{A minimal model of dynamical phase transition},  {\href{https://iopscience.iop.org/article/10.1209/0295-5075/116/50009}{Europhys. Lett. \textbf{116}, 50009 (2016).}} 
\bibitem{NT2018DPT} P. Tsobgni Nyawo and H. Touchette, \textit{Dynamical phase transition in drifted Brownian motion} {\href{https://journals.aps.org/pre/abstract/10.1103/PhysRevE.98.052103}{Phys. Rev. E \textbf{98}, 052103, (2018)}}


\bibitem{exclusion} T. Bodineau, B. Derrida, \textit{Distribution of current in nonequilibrium diffusive systems and phase transitions}, {\href{https://journals.aps.org/pre/abstract/10.1103/PhysRevE.72.066110} {Phys. Rev. E
\textbf{72}, 066110 (2005)}.}
\bibitem{glass} J. P. Garrahan, R. L. Jack, V. Lecomte, E. Pitard, K. van Duijvendijk, F. van Wijland,     \textit{Dynamical First-Order Phase Transition in Kinetically Constrained Models of Glasses}, {\href{https://journals.aps.org/prl/abstract/10.1103/PhysRevLett.98.195702} {Phys. Rev. Lett. \textbf{98}, 195702 (2007)}.}
\bibitem{kafri} G. Bunin, Y. Kafri, D. Podolosky, \textit{Non-differentiable large-deviation functionals in boundary-driven diffusive systems}, {\href{https://iopscience.iop.org/article/10.1088/1742-5468/2012/10/L10001} {J. Stat. Mech. \textbf{2012}, L10001, (2012)}.}
\bibitem{exclusion1} C. P. Espigares, P. L. Garrido,  P. I. Hurtado, \textit{Dynamical
phase transition for current statistics in a simple driven diffusive
system}, {\href{https://journals.aps.org/pre/abstract/10.1103/PhysRevE.87.032115} {Phys. Rev. E \textbf{87}, 032115 (2013)}.}
\bibitem{singularities}  Y. Baek, Y. Kafri, \textit{Singularities in large deviation functions}, {\href{https://iopscience.iop.org/article/10.1088/1742-5468/2015/08/P08026/pdf} {J. Stat. Mech. \textbf{2015}, P08026 (2015)}.} 

\bibitem{baek} Y. Baek, Y. Kafri, V. Lecomte, \textit{Dynamical Symmetry Breaking and Phase Transitions in Driven Diffusive Systems},  {\href{https://journals.aps.org/prl/abstract/10.1103/PhysRevLett.118.030604} { Phys. Rev. Lett. \textbf{118}, 030604 (2017) }.}
\bibitem{baek1} Y. Baek, Y. Kafri, V. Lecomte, \textit{Dynamical phase transitions in the current distribution of driven diffusive channels},  {\href{https://iopscience.iop.org/article/10.1088/1751-8121/aaa8f9} { J. Phys. A: Math. Theor. \textbf{51} 105001 (2018) }.}

\bibitem{MukherjeeSmith23} S. Mukherjee and N. R. Smith, {\href{https://journals.aps.org/pre/abstract/10.1103/PhysRevE.107.064133} {Phys. Rev. E \textbf{107}, 064133, (2023).}}

\bibitem{Smith22Chaos} N. R. Smith, \textit{Large deviations in chaotic systems: Exact results and dynamical phase transition}, {\href{https://journals.aps.org/pre/abstract/10.1103/PhysRevE.106.L042202}{Phys. Rev. E \textbf{106}, L042202, (2022)}}. 

\bibitem{MukherjeeSmithConvexhull} S. Mukherjee, N. R. Smith, \textit{Large deviations in statistics of the convex hull of passive and active particles: A theoretical study}, {\href{
https://journals.aps.org/pre/abstract/10.1103/PhysRevE.109.044120
}{	Phys. Rev. E \textbf{109}, 044120, (2024)}}


\bibitem{MBE19} E. Mallmin, R. A. Blythe and M. R. Evans, \textit{A comparison of dynamical fluctuations of biased diffusion and run-and-tumble dynamics in one dimension}, {\href{https://iopscience.iop.org/article/10.1088/1751-8121/ab4349}{J. Phys. A: Math. Theor. \textbf{52} 425002 (2019)}}.


\bibitem{Weiss84} G. H. Weiss, \textit{First passage times for correlated random walks and some generalizations}, {\href{https://link.springer.com/article/10.1007/BF01011837} { J. Stat. Phys. \textbf{37}, 325 (1984) }}.

\bibitem{WMLW87} G. H. Weiss, J. Masoliver, K. Lindenberg, and B. J. West, \textit{First-passage times for non-Markovian processes: Multivalued noise}, {\href{https://journals.aps.org/pra/abstract/10.1103/PhysRevA.36.1435} {Phys. Rev. A \textbf{36}, 1435 (1987).}} 

\bibitem{MPW92} J. Masoliver, J. M. Porrà, and G. H. Weiss, \textit{Solutions of the telegrapher’s equation in the presence of traps}, {\href{https://journals.aps.org/pra/abstract/10.1103/PhysRevA.45.2222} {Phys. Rev. A \textbf{45}, 2222 (1992)}}. 


\bibitem{MalakarEtAl18}
K. Malakar, V. Jemseena, A. Kundu, K. Vijay Kumar, S. Sabhapandit, S. N. Majumdar, S. Redner and A. Dhar, \textit{Steady state, relaxation and first-passage properties of a run-and-tumble particle in one-dimension}, {\href{https://iopscience.iop.org/article/10.1088/1742-5468/aab84f}{J. Stat. Mech. \textbf{2018} 043215 (2018)}}.

\bibitem{MLMS20} F. Mori, P. Le Doussal, S. N. Majumdar, and G. Schehr, \textit{Universal Survival Probability for a $d$-Dimensional Run-and-Tumble Particle}, {\href{https://journals.aps.org/prl/abstract/10.1103/PhysRevLett.124.090603} {Phys. Rev. Lett. \textbf{124}, 090603 (2020)}}; \textit{Universal properties of a run-and-tumble particle in arbitrary dimension}, {\href{https://journals.aps.org/pre/abstract/10.1103/PhysRevE.102.042133}{Phys. Rev. E \textbf{102}, 042133 (2020)}}.


 \bibitem{BMS21} B. De Bruyne, S. N. Majumdar and G. Schehr, \textit{Survival probability of a run-and-tumble particle in the presence of a drift}, {\href{https://iopscience.iop.org/article/10.1088/1742-5468/abf5d5/meta} {J. Stat. Mech. \textbf{2021} 043211 (2021)}}


\bibitem{NS24} S. K. Nath, S. Sabhapandit, \textit{Survival probability and position distribution of a run and tumble particle in $U(x)=\alpha|x|$ potential with an absorbing boundary}, arXiv:2405.18988


\bibitem{HatanoNelson96}
N. Hatano, D.R. Nelson, \textit{Localization transitions in non-Hermitian quantum mechanics}. {\href{https://journals.aps.org/prl/abstract/10.1103/PhysRevLett.77.570} {Phys. Rev. Lett., \textbf{77(3)}, 570 (1996)}}.

\bibitem{TC08} J. Tailleur, M. E. Cates, \textit{Statistical Mechanics of Interacting Run-and-Tumble Bacteria}, {\href{https://journals.aps.org/prl/abstract/10.1103/PhysRevLett.100.218103}{Phys. Rev. Lett. \textbf{100}, 218103 (2008)}}; \textit{Sedimentation, trapping, and rectification of dilute bacteria}, {\href{https://iopscience.iop.org/article/10.1209/0295-5075/86/60002}{Europhys. Lett. \textbf{86}, 60002 (2009)}}.

\bibitem{footnote:conditional}
Since  $P_{\pm}(x,A,t)$ must be positive, so are  $G_{k \pm} (x,t)$,
hence from \eqref{asympt} it is possible to choose the (unnormalized) eigenfunctions $\Psi_{k\pm} (x)$
to be positive, which then implies that the even eigenfunction must be $\Phi_1(x)$. Furthermore, with this choice, these functions give information on the PDF of the final positions conditioned to the value of $a=A/T$, i.e.
on the "optimal configuration" associated to the large deviations, see \cite{CH15} for analogous arguments in the case of Brownian motion.


\bibitem{CH15} R. Chetrite and H. Touchette, \textit{Nonequilibrium Markov Processes Conditioned on Large Deviations}, {\href{https://link.springer.com/article/10.1007/s00023-014-0375-8}{Ann. Henri Poincare \textbf{16}, 2005 (2015)}}.

%
\bibitem{MM22} A. Mazzolo and C. Monthus, \textit{Conditioning diffusion processes with respect to the local time at the origin}, {\href{https://iopscience.iop.org/article/10.1088/1742-5468/ac9618}{J. Stat. Mech. (2022) 103207}}.

\bibitem{HNE16} J. Hoppenau, D. Nickelsen and A. Engel, \textit{Level 2 and level 2.5 large deviation functionals for systems with and without detailed balance}, {\href{https://iopscience.iop.org/article/10.1088/1367-2630/18/8/083010}{New J. Phys. \textbf{18} 083010 (2016)}}.


\bibitem{CVC22} G. Carugno, P. Vivo and F. Coghi, \textit{Graph-combinatorial approach for large deviations of Markov chains}, {\href{https://iopscience.iop.org/article/10.1088/1751-8121/ac79e6}{J. Phys. A: Math. Theor. \textbf{55} 295001 (2022)}}.

\bibitem{Monthus24} C. Monthus, \textit{Large deviations at level 2.5 and for trajectories observables of diffusion processes: the missing parts with respect to their random-walks counterparts}, {\href{https://iopscience.iop.org/article/10.1088/1751-8121/ad26ae}{J. Phys. A: Math. Theor. \textbf{57} 095002 (2024)}}.


\bibitem{BLLRVV2016} C. Bechinger, R. Di Leonardo, H. Löwen, C. Reichhardt, G. Volpe, and G. Volpe, \textit{Active particles in complex and crowded environments}, {\href{https://journals.aps.org/rmp/abstract/10.1103/RevModPhys.88.045006}{Rev. Mod. Phys. \textbf{88}, 045006, (2016). }}

\bibitem{LMS19} P. Le Doussal, S. N. Majumdar, and G. Schehr, \textit{Noncrossing run-and-tumble particles on a line}, {\href{https://journals.aps.org/pre/abstract/10.1103/PhysRevE.100.012113} {Phys. Rev. E \textbf{100}, 012113 (2019)}}.


\bibitem{PBDN21} A. Poncet, O. Bénichou, V. Démery, and D. Nishiguchi, \textit{Pair correlation of dilute active Brownian particles: From low-activity dipolar correction to high-activity algebraic depletion wings}, {\href{https://journals.aps.org/pre/abstract/10.1103/PhysRevE.103.012605}{Phys. Rev. E \textbf{103}, 012605 (2021)}}. 

\bibitem{RSBI22} P. Rizkallah, A. Sarracino, O. Bénichou, and P. Illien, \textit{Microscopic Theory for the Diffusion of an Active Particle in a Crowded Environment}, {\href{https://journals.aps.org/prl/abstract/10.1103/PhysRevLett.128.038001} {Phys. Rev. Lett. \textbf{128}, 038001 (2022).}} 

\bibitem{LMS21} P. Le Doussal, S. N. Majumdar, and G. Schehr, \textit{Stationary nonequilibrium bound state of a pair of run and tumble particles}, {\href{https://journals.aps.org/pre/abstract/10.1103/PhysRevE.104.044103}{Phys. Rev. E \textbf{104}, 044103 (2021)}}. 

\bibitem{Singh21}
P. Singh, A. Kundu, \textit{Crossover behaviours exhibited by fluctuations and correlations in a chain of active particles}, {\href{https://iopscience.iop.org/article/10.1088/1751-8121/ac0a9f}{J. Phys. A: Math. Theor. \textbf{54}, 305001 (2021)}}.

\bibitem{ARYL21} T. Agranov, S. Ro, Y. Kafri, and V. Lecomte, \textit{Exact fluctuating hydrodynamics of active lattice gases -- typical fluctuations}, {\href{https://iopscience.iop.org/article/10.1088/1742-5468/ac1406}{J. Stat. Mech \textbf{2021}, 083208 (2021)}}.

\bibitem{Cates22} T. Banerjee, R. L. Jack and M. E. Cates, \textit{Tracer dynamics in one dimensional gases of active or passive particles}, {\href{https://iopscience.iop.org/article/10.1088/1742-5468/ac4801/meta}{J. Stat. Mech. \textbf{2022} 013209 (2022)}}.

\bibitem{ARYL22} T. Agranov, S. Ro, Y. Kafri, and V. Lecomte, \textit{Macroscopic Fluctuation Theory and current fluctuations in active lattice gases}, {\href{https://scipost.org/SciPostPhys.14.3.045}{SciPost Phys. \textbf{14}, 045 (2023)}}.

\bibitem{KPS23} V. Kumar, A. Pal, O. Shpielberg, \textit{Arrhenius law for interacting diffusive systems}, {\href{https://journals.aps.org/pre/abstract/10.1103/PhysRevE.109.L032101}{Phys. Rev. E \textbf{109}, L032101 (2024)}} .
\bibitem{footnote:N} Technically, the RTP starts at $x_0=\epsilon$ with $\epsilon \to 0^+$, and one does not count the possible first crossing at time $t=\epsilon/v_0$.


\bibitem{BesselAsymptoticsDLMF} \url{https://dlmf.nist.gov/10.41}




\end{thebibliography}
\end{document}